\documentclass[fleqn,usenatbib]{mnras}

\usepackage{mathptmx}

\usepackage[T1]{fontenc}

\DeclareRobustCommand{\VAN}[3]{#2}
\let\VANthebibliography\thebibliography
\def\thebibliography{\DeclareRobustCommand{\VAN}[3]{##3}\VANthebibliography}

\usepackage{graphicx}	
\usepackage{amsmath}	
\usepackage{amssymb}	
\usepackage{multirow}
\usepackage{hyperref}

\newcommand{\DSPSR}{\texttt{DSPSR}}
\newcommand{\PRESTO}{\texttt{PRESTO}}

\newcommand{\PSRCHIVE}{\texttt{PSRCHIVE}}
\newcommand{\TEMPO}{\texttt{TEMPO}}

\newcommand{\PULSARMINER}{\texttt{PULSAR\_MINER}}
\newcommand{\SPIDERTWISTER}{\texttt{SPIDER\_TWISTER}}

\newcommand{\degree}{~{\rm deg}}

\newcommand{\dmunit}{pc\,cm$^{-3}$}
\newcommand{\pmunit}{mas\,yr$^{-1}$}

\newcommand{\msun}{M$_{\sun}$}

\newcommand{\us}{$\mu$s}

\newcommand{\Tasc}{T_{\rm asc}}

\newcommand{\Pobs}{P_{\rm obs}}
\newcommand{\Pdot}{\dot{P}}
\newcommand{\Pbdotobs}{\dot{P_{\rm b}}_{\rm ,obs}}
\newcommand{\Pdotobs}{\dot{P}_{\rm obs}}
\newcommand{\Pdotint}{\dot{P}_{\rm int}}
\newcommand{\Pint}{P_{\rm int}}

\newcommand{\Pb}{P_{\rm b}}
\newcommand{\phib}{\phi_{\rm b}}

\newcommand{\fc}{f_{\rm c}}

\newcommand{\Mp}{M_{\rm p}}
\newcommand{\Mc}{M_{\rm c}}
\newcommand{\Mtot}{M_{\rm tot}}

\newcommand{\xp}{x_{\rm p}}

\newcommand{\omegadot}{\dot{\omega}}

\newcommand{\zmax}{z_{\rm max}}
\newcommand{\rcore}{r_{\rm c}}
\newcommand{\rhalflight}{r_{\rm hl}}
\newcommand{\BWeff}{{\rm BW}_{\rm eff}}
\newcommand{\h}{^{\rm h}}
\newcommand{\m}{^{\rm m}}
\newcommand{\s}{^{\rm s}}

\title[The MeerKAT globular cluster census]{Eight new millisecond pulsars from the first MeerKAT globular cluster census}

\author[A. Ridolfi et al.]{\parbox{\textwidth}{A.~Ridolfi,$^{1,2}$\thanks{E-mail: alessandro.ridolfi@inaf.it}
T.~Gautam,$^{2}$\thanks{E-mail: tgautam@mpifr-bonn.mpg.de}
P.~C.~C.~Freire,$^{2}$
S.~M.~Ransom,$^{3}$
S.~J.~Buchner,$^{4}$
A.~Possenti,$^{1,5}$
V.~Venkatraman~Krishnan,$^{2}$
M.~Bailes,$^{6,7}$
M.~Kramer,$^{2,8}$
B.~W.~Stappers,$^{8}$
F.~Abbate,$^{2}$
E.~D.~Barr,$^{2}$
M.~Burgay,$^{1}$
F.~Camilo,$^{4}$
A.~Corongiu,$^{1}$
A.~Jameson,$^{6,7}$
P.~V.~Padmanabh,$^{2}$
L.~Vleeschower,$^{8}$
D.~J.~Champion,$^{2}$
M.~Geyer,$^{4}$
A.~Karastergiou,$^{9,10}$
R.~Karuppusamy,$^{2}$
A. Parthasarathy,$^{2}$
D.~J.~Reardon,$^{6,7}$
M.~Serylak,$^{4}$
R.~M.~Shannon$^{6,7}$ and
R.~Spiewak$^{8,6}$
} \\ \\ \\ \\
\parbox{\textwidth}{
$^1$ INAF -- Osservatorio Astronomico di Cagliari, Via della Scienza 5, I-09047 Selargius (CA), Italy\\
$^2$ Max-Planck-Institut f\"{u}r Radioastronomie, Auf dem H\"{u}gel 69, D-53121 Bonn, Germany\\
$^{3}$ National Radio Astronomy Observatory, 520 Edgemont Rd., Charlottesville, VA 22903, USA\\
$^{4}$ South African Radio Astronomy Observatory (SARAO), 2 Fir Street, Black River Park, Observatory, Cape Town, 7925, South Africa\\
$^{5}$ Universit\`a di Cagliari, Dipartimento di Fisica,  S.P. Monserrato-Sestu Km 0,700,  I-09042 Monserrato (CA), Italy \\
$^{6}$  Centre for Astrophysics and Supercomputing, Swinburne University of Technology, P.O. Box 218, Hawthorn, VIC 3122, Australia\\
$^{7}$ ARC Centre of Excellence for Gravitational Wave Discovery (OzGrav)\\
$^{8}$Jodrell Bank Centre for Astrophysics, Department of Physics and Astronomy, The University of Manchester, Manchester M13 9PL, UK\\
$^{9}$Department of Astrophysics, University of Oxford, Denys Wilkinson Building, Keble Road, Oxford OX1 3RH, UK\\
$^{10}$Department of Physics and Electronics, Rhodes University, PO Box 94, Grahamstown 6140, South Africa\\
}
}

\date{Accepted XXX. Received YYY; in original form ZZZ}

\pubyear{2020}

\begin{document}
\label{firstpage}
\pagerange{\pageref{firstpage}--\pageref{lastpage}}
\maketitle

\begin{abstract}
We have used the central 44 antennas of the new 64-dish MeerKAT radio telescope array to conduct a deep search for new pulsars in the core of nine globular clusters. This has led to the discovery of eight new millisecond pulsars in six different clusters.
Two new binaries, 47~Tuc~ac and 47~Tuc~ad, are eclipsing ``spiders'', featuring compact orbits ($\lesssim 0.32$ days), very low-mass companions and regular occultations of their pulsed emission. The other three new binary pulsars (NGC 6624G, M62G, and Ter 5 an) are in wider ($> 0.7$ days) orbits, with companions that are likely to be white dwarfs or neutron stars.  NGC 6624G has a large eccentricity of $e\simeq 0.38$, which enabled us to detect the rate of advance of periastron. This suggests that the system is massive, with a total mass of $\Mtot = 2.65 \pm 0.07$~\msun. Likewise, for Ter 5 an, with $e \simeq 0.0066$, we obtain $\Mtot= 2.97 \pm 0.52$~\msun.
The other three new discoveries (NGC~6522D, NGC~6624H and NGC~6752F) are faint isolated pulsars. 
Finally, we have used the whole MeerKAT array and  synthesized 288 beams, covering an area of $\sim2$~arcmin in radius around the center of NGC 6624. This has allowed us to localize many of the pulsars in the cluster, demonstrating the beamforming capabilities of the TRAPUM software backend and paving the way for the upcoming MeerKAT globular cluster pulsar survey. 
\end{abstract}



\section{Introduction}
\label{sec:intro}
Globular clusters (GCs) are renowned to be among the most fertile grounds for the formation of millisecond pulsars (MSPs). Boasting stellar densities that exceed those found in the Galactic disk by several orders of magnitude, GCs promote two- and three-body gravitational interactions, in the form of tidal captures, exchange encounters and binary disruptions \citep{Hills1975,Sigurdsson_Phinney1995}. As a result, isolated neutron stars (NSs) can form binaries with other stars, most often a main sequence or giant star. The latter then evolves and eventually spins the NS up through the transfer of mass and angular momentum \citep{Alpar+1982,Radhakrishnan_Srinivasan1982,Bhattacharya_vandenHeuvel1991}, a phase in which the system is is seen as a ``low mass X-ray binary'' (LMXB). This phase can last hundreds of millions to billions of years. At the end of the process, the magnetic field of the NS has been ablated, and the object spins hundreds of times per second.

In the Galactic disk, LMXBs can only evolve from binaries where the
NS stays bound to a companion star after the supernova event that forms it. This happens only to a small minority of NSs, all others, seen as isolated rotation-powered NSs, eventually become undetectable as they spin down over time. In GCs, however, the dominant formation channel for LMXBs are exchange encounters  involving such isolated, undetectable NSs. This is the reason why, per unit of stellar mass, GCs have three orders of magnitude more LMXBs than the Galactic disk. These NSs can thus be resurrected through accretion, and when the latter stops, the LMXB becomes a binary MSP. The ultimate proof of this scenario was provided a few years ago by the ``transitional'' binary pulsar PSR J1824$-$2452I, located in the GC M28, which showed swings between LMXB and radio-MSP phases over timescales as short as just a few weeks \citep{Papitto+2013}. 

The dynamic environments of GCs can also create exotic binary pulsars that cannot form in the Galaxy.
One of the mechanisms for this are the ``secondary'' exchange encounters, where the NS involved is a MSP that has already been recycled as a consequence of a previous exchange encounter. These are more likely to occur in the GCs with the densest cores \citep{Verbunt_Freire_14}, which can produce highly eccentric binary MSPs with massive companions, possibly other NSs. Examples are M15C \citep{Prince1991},
NGC 1851A \citep{Freire+2004,Ridolfi+2019}, NGC~6544B \citep{Lynch+2012} and NGC~6652A \citep{DeCesar+2015}. Such secondary exchange encounters could potentially produce MSP--black hole and even MSP--MSP binaries \citep[e.g.][]{Ransom+2008}, which would open up unprecedented possibilities for fundamental physics experiments (e.g. \citealt{Liu+2014}).
Some of these secondary exchange encounters could also place an already recycled pulsar in orbit with a main sequence star, resulting eventually in a new LMXB system, where it undergoes further recycling. This is a possible explanation for the many fast-spinning pulsars in Terzan~5 (henceforth, Ter 5), which include the fastest-spinning pulsar known, Ter~5~ad \citep{Hessels+2006}.

Even the usually less exciting isolated pulsars have proven to be extremely valuable, when found in GCs. Timing of single objects \citep{Perera+2017} or of an ensemble of pulsars \citep{Freire+2017,Prager+2017,Abbate+2018,Abbate+2019} has been exploited to constrain the gravitational potential of several GCs and can indicate the possible presence of an intermediate-mass black hole at their center. The measurement of the dispersion measures (DMs) of the pulsars in 47 Tucanae (hereafter, 47 Tuc) has allowed the first detection of an intra-cluster ionized gas, enabling the three-dimensional localization of the pulsars in the cluster \citep{Freire+2001b,Abbate+2018}. Furthermore, the study of the polarimetric properties of the same pulsars has made it possible to study the Galactic magnetic field at arcsecond scales for the very first time, in turn revealing the existence of an extended magnetized outflow from our Galaxy \citep{Abbate+2020a}.

Since the discovery of the first pulsar in a GC by \citet{Lyne+1987}, all of these breakthroughs have nourished interest in GC pulsars, and sustained an ever-increasing effort in the search for these objects. 
This has led to the discovery of over two hundred pulsars in at least 36 different clusters\footnote{See \url{https://www3.mpifr-bonn.mpg.de/staff/pfreire/GCpsr.html} for an up-to-date list of the known pulsars in GCs.}. 
However, after a burst of new discoveries made in the early 2000's, only a few new GC pulsars were found in the next several years. Further discoveries were essentially hampered by the sensitivity limit reached at the largest available telescopes, namely Arecibo and the Green Bank Telescope (GBT) for the northern GCs, and the Parkes radio telescope for the southern GCs. New pulsars were mostly found through the reprocessing of archival data using new search techniques, such as the stacking of Fourier spectra from multiple epochs \citep{Pan+2016,Cadelano+2018} or ``jerk'' searches \citep{Andersen_Ransom2018}. 
This situation has greatly changed since the mid 2010's. New, wide-bandwidth receivers as well as more modern digital backends have recently been installed at the Upgraded Giant Metrewave Radio Telescope \citep[uGMRT,][]{Gupta+2017} and at Parkes \citep{Hobbs+2020}. These have already led to the discovery of a new steep-spectrum MSP in NGC 6652 (Gautam et al., in prep.) as well as five faint MSPs in $\omega$~Centauri \citep{Dai+2020}.
Moreover, entirely new telescopes have been built in both hemispheres. Since 2016, the new Chinese Five-hundred-meter Aperture Spherical Telescope \citep[FAST,][]{Nan+2011} has been providing a factor of 2 to 3 times better raw sensitivity than Arecibo in the $-15^\circ$ to $+65^\circ$ declination range of the sky. Early FAST GC observations have resulted, at the time of writing, in the discovery of 32 pulsars\footnote{See also \url{https://fast.bao.ac.cn/cms/article/112}} (\citealt{Wang+2020}, \citealt{Pan+2020}).

The year 2018 also saw the inauguration of the South African 64-dish MeerKAT radio telescope array \citep{Booth_Jonas2012}, the precursor of the Square Kilometer Array - SKA1-mid \citep{Dewdney+2009}. As it is located in the Karoo desert at a latitude of $-30^\circ$, MeerKAT is the only radio telescope, other than Parkes, to have access to high-sensitivity, cm-wavelength observations of pulsars in GCs with declinations of $\delta \lesssim-45^\circ$. When using all the 64 dishes, MeerKAT boasts a gain of 2.8 K\,Jy$^{-1}$, four times higher than that of Parkes and 1.4 times higher than that of the GBT. Such an improvement in raw sensitivity represents a major leap for studying pulsars in southern GCs. 

Two MeerKAT Large Survey Projects (LSPs) that include science of pulsars in GCs as part of their main scientific goals have already been approved and commenced.
The first one to start was MeerTime\footnote{\url{http://www.meertime.org}} \citep{Bailes+2020}, which began collecting data in early 2019. The project has a variety of scientific objectives, one of these is the exploitation of GC pulsars through pulsar timing and polarimetry. Using the Pulsar Timing User Supplied Equipment (PTUSE) machines as the main data acquisition system, MeerTime observations can record the signal from up to four (but only one was available before December 2019) tied-array beams in either ``timing'' or ``search'' mode\footnote{In ``search'' mode the observing band is divided into several small channels and recorded every few $\mu$s onto a ``filterbank'' file type. In ``timing'' mode, the signal of a particular pulsar is folded in real time according to a model (the ephemeris) that describes its rotational behaviour, resulting in a ``folded archive'' file.}, with very high time resolution, full polarimetric information, and real-time coherent de-dispersion. These characteristics are an asset to multiple scientific applications for objects within the diameter of the tied array beam. For example, one cannot only fold one known pulsar in a GC, but also look for giant pulses from other pulsars in the cluster, as well as search for new pulsars in the same data.
The search and discovery of new extreme and exotic pulsars in GCs is one of the main objectives of the second LSP, dubbed TRAPUM\footnote{\url{http://www.trapum.org}} (TRAnsients and PUlsars with MeerKAT, \citealt{Stappers_Kramer2016}). Since April 2020, the TRAPUM project has been making use of the full MeerKAT array to observe several GCs. A 60-node computing cluster (APSUSE) utilizes beamforming techniques to synthesize and record incoherently de-dispersed search-mode data for up to $\sim400$ beams on the sky. The latter cover an area of $\sim 1-4$ arcmin in radius, depending on the number of antennas used, the observing frequency, and the beam overlap. This allows the search for pulsars not only in the core, but also in the outskirts of the observed globulars, with full sensitivity.
Given the complementary scientific goals and characteristics of their data acquisition systems, MeerTime and TRAPUM LSPs decided to collaborate on a joint endeavour to time and search for pulsars in GCs. 
This has already resulted in the most detailed study of the giant pulses from PSR J1823$-$3021A in NGC 6624 \citep{Abbate+2020b}.

Here we report on the results of the first GC census observations conducted with MeerKAT in the framework of the joint MeerTime \& TRAPUM activities. They were mostly carried out using the 44 central antennas of the array (although only up to 42 were available at once) and the MeerTime PTUSE data acquisition system, while TRAPUM’s backend has been used to quickly ascertain the positions of some of pulsars within their GC. Besides producing new scientific results, the census has served as a benchmark upon which to tune and execute the TRAPUM experiment, which will find pulsars with full sensitivity, both within and beyond the GC core.

In Section \ref{sec:observations} we report on the target selection, the instrumental set-up, and the observations. In Section \ref{sec:analysis} we describe the data reduction and analysis methods. The main results are detailed in Section \ref{sec:results}, and their implications are discussed in Section \ref{sec:discussion}. Finally, in Section \ref{sec:conclusions}, we draw some conclusions and comment on future prospects of GC pulsar observations with MeerKAT.

\begin{table*}
\caption[]{List of globular clusters observed in the context of this work, ordered by increasing Right Ascension, and with their basic parameters as reported by \citet{Harris2010}, unless otherwise specified. $\rcore$: core radius; $\rhalflight$: half-light radius; $\langle {\rm DM} \rangle$: median DM of the known pulsars; $N_{\rm p, isol}$: number of isolated pulsars; $N_{\rm p, bin}$: number of binary pulsars.}
\label{tab:observed_clusters}
\footnotesize
\centering
\renewcommand{\arraystretch}{1.0}
\vskip 0.1cm
\begin{tabular}{lcccccccc}
\hline
\hline
\multicolumn{9}{c}{Observed Clusters}\\
\hline
    Cluster     & Center coordinates &  Distance & Core  & $\rcore$  & $\rhalflight$ & $\langle {\rm DM} \rangle$  &  $N_{\rm p, isol}$    & $N_{\rm p, bin}$  \\
 name          & (RA, Dec)   & (kpc)   & collapsed?  & (arcmin) &  (arcmin) & (\dmunit)                  & (old / new) & (old / new) \\
 \hline
NGC 104 (47 Tuc)& 00$\h$\,24$\m$\,05.67$\s$, $-72\degr\,04\arcmin\,52.6\arcsec$ &  4.69$^{\rm a}$   & No       & 0.36    & 3.17 & \phantom{0}24.4 &   10 / --     & 15 / 2     \\ 
NGC 6266 (M62)  & 17$\h$\,01$\m$\,12.80$\s$, $-30\degr\,06\arcmin\,49.4\arcsec$ &  6.80   & No      & 0.22    & 0.92 & 114.0           &   0 / --    &   6 / 1     \\
NGC 6397        & 17$\h$\,40$\m$\,42.09$\s$, $-53\degr\,40\arcmin\,27.6\arcsec$ &  2.30   & Yes      & 0.05    & 2.90 & \phantom{0}71.8 &  0 / --   &  1 / --  \\
Terzan 5        & 17$\h$\,48$\m$\,04.80$\s$, $-24\degr\,46\arcmin\,45.0\arcsec$ & 6.90    & No       & 0.16    & 0.72 & 237.9 &  19 / --    &   19 / 1 \\
NGC 6522        & 18$\h$\,03$\m$\,34.02$\s$, $-30\degr\,02\arcmin\,02.3\arcsec$ & 7.70    & Yes      & 0.24    & 1.00 & 192.7 &  3 / 1    &   0 / - \\
NGC 6624        & 18$\h$\,23$\m$\,40.51$\s$, $-30\degr\,21\arcmin\,39.7\arcsec$ &  7.90   & Yes      & 0.06    & 0.82 & \phantom{0}86.9 &  5 / 1    &  1 / 1    \\
NGC 6626 (M28)  & 18$\h$\,24$\m$\,32.81$\s$, $-24\degr\,52\arcmin\,11.2\arcsec$ & 5.50    & No       & 0.05    & 1.97 & 119.7 &  4 / --    &  8 / --   \\
NGC 6752        & 19$\h$\,10$\m$\,52.11$\s$, $-59\degr\,59\arcmin\,04.4\arcsec$  &  4.00   & Yes      & 0.17    & 1.91 & \phantom{0}33.3 &  4 / 1     &  1 / --    \\
NGC 7099 (M30)  & 21$\h$\,40$\m$\,22.12$\s$, $-23\degr\,10\arcmin\,47.5\arcsec$ & 8.10    & Yes      & 0.06    & 1.03 & \phantom{0}25.1 &  0 / --     & 2 / -- \\
\hline
\multicolumn{9}{l}{$^{\rm a}$ Taken from \citet{Woodley+2012}.}
\end{tabular}
\end{table*}


\section{Observations}
\label{sec:observations}

\subsection{Target selection}
\label{sec:selection}

\begin{figure}

\centering
	\includegraphics[width=\columnwidth]{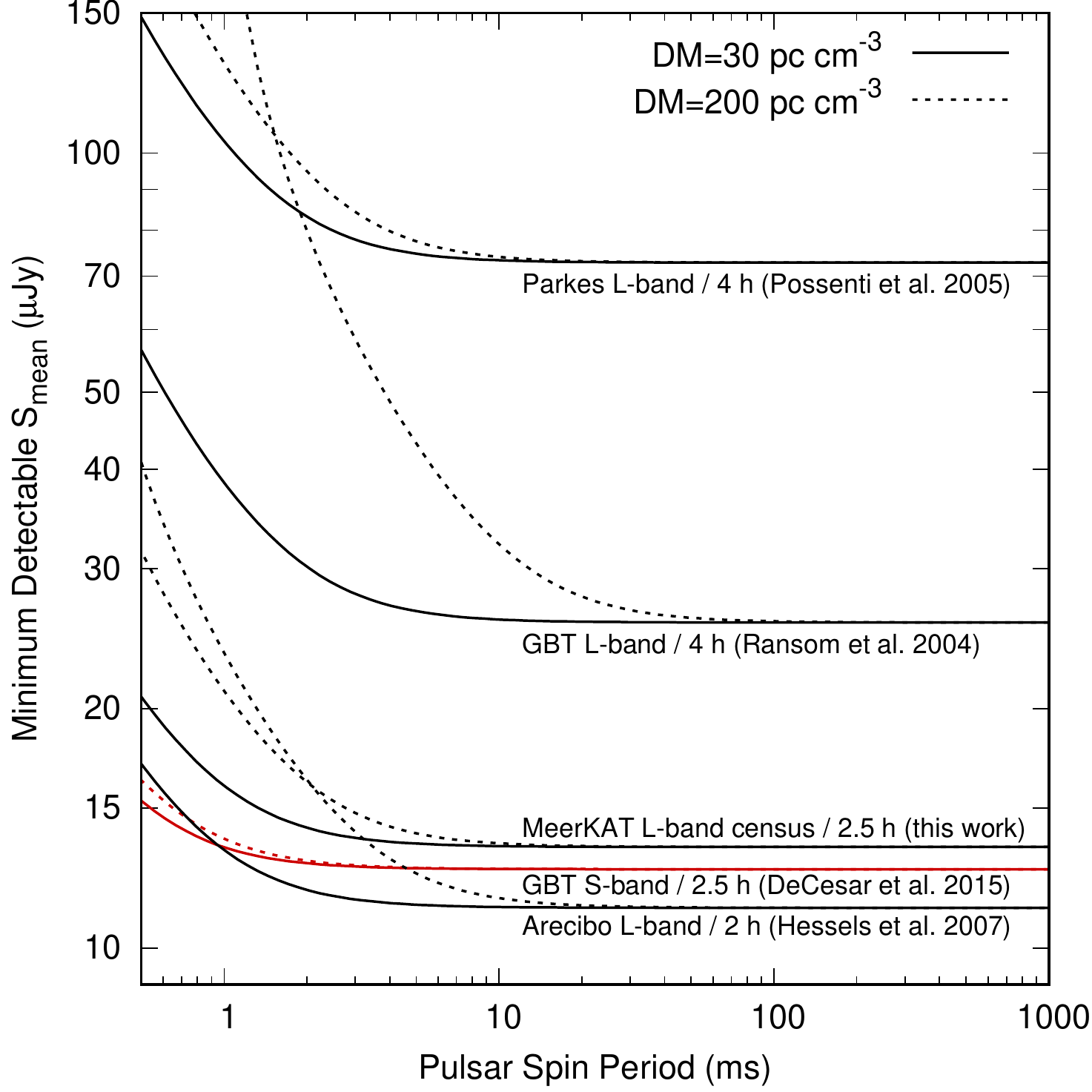}
  	\caption{Sensitivity curves of the major L-band (black) and S-band (red) GC pulsar surveys conducted with the Parkes, Arecibo, GBT and MeerKAT telescopes. The curves are calculated  using the survey parameters reported in the references, the typical integration time as indicated in the labels, and assuming an elevation of $45\degree$.  Solid lines are relative to a ${\rm DM}=30$~\dmunit, dashed lines to a ${\rm DM}~=~200$~\dmunit. Note (1): The GBT S-band survey appears to be more sensitive because we are not correcting for the spectral index. Its sensitivity to pulsars is very similar to that of the GBT L-band survey, since both surveys detect very similar numbers of pulsars in the same  clusters. Note (2): Other GC pulsar surveys, not reported in this plot, are currently being conducted with more modern telescopes and instrumentation (e.g. \citealt{Pan+2020} and \citealt{Dai+2020})}.
  	\label{fig:sensitivity}

\end{figure}

Among all the GCs visible from the MeerKAT site, we selected a sample of nine. These are listed, along with their main characteristics, in Table~\ref{tab:observed_clusters}.
The clusters were chosen based on three main criteria:
\begin{itemize}
\item Each of these clusters hosts at least one previously known pulsar. This requirement had a twofold purpose: firstly, the DM of each cluster is constrained within a relatively narrow range, greatly easing the search for new pulsars; secondly, the re-detection of the known pulsars allows performance comparisons against other telescopes and serves as a testbed for the data reduction and searching pipelines used.
\item Among the GCs with known pulsars, we prioritized those farther South, which had only been searched with the Parkes telescope. For these, the sensitivity gains from MeerKAT searches are larger (see Figure~\ref{fig:sensitivity}).
\item Five of the clusters we targeted - the majority of the sample - are core-collapsed clusters. The reasons for this choice have been highlighted in the Introduction: these are the clusters where exotic binary systems containing MSPs and massive, compact companions (including MSP--black hole, or MSP--MSP systems) are likely to form.
\end{itemize}
However, not all the selected clusters are core-collapsed: Ter~5 and 47~Tuc are so rich that they offer a good likelihood of new discoveries with improved sensitivity. Furthermore, searching a diversity of types of clusters is important for improving our understanding of the pulsar populations of different GCs, and in particular their differences.

The chosen clusters also have a range of DMs, allowing us to test the telescope performance and the data analysis strategies in a number of different cases. The pulsars in the clusters with a low DM ($<50$~\dmunit) scintillate over timescales of just a few hours (e.g. 47~Tuc), whereas those in clusters with a high DM ($> 150$~\dmunit) show fairly stable flux densities (e.g. NGC~6522). This results in different search strategies: while the former are targeted with short, but frequent integrations, the latter are targeted with deeper, but less frequent integrations.

\begin{table*}
\renewcommand{\arraystretch}{1.0}
\setlength{\tabcolsep}{0.15cm}
\footnotesize
\caption{List of MeerKAT observations made in the context of this work. All the observations, with the exception of the TRAPUM ones, were taken using the 1-km core antennas and the search-mode data were acquired using the PTUSE pulsar processor with full-Stokes, a time resolution of 9.57~$\mu$s, and coherently de-dispersing at the nominal cluster DM. TRAPUM observations (marked by ``-trapum'' in their observation id) were taken using 60 dishes and acquiring search-mode data with a 76~$\mu$s time resolution  and 4096 frequency channels, without coherent de-dispersion. Observations that were searched for new pulsars as outlined in Section \ref{sec:analysis} are marked with a star ($^\star$); first detections of new pulsars (i.e. discoveries) are highlighted in bold.}
\label{tab:list_observations}
\begin{tabular}{ccccccccc}
\hline
Cluster          & Obs.              & Start Epoch                    & $\Delta t_{\rm obs}$ & $f_{\rm c}$  & BW    & $\Delta f_{\rm ch} $ & Redetections of & Detections   \\
                 &  id                   &  (MJD)                      & (s)               & (MHz)        & (MHz) & (MHz)           & previously known pulsars     & of new pulsars \\
\hline
47 Tuc   & \multirow{2}{*}{01L$^{\star}$}  & \multirow{2}{*}{58557.386} & \multirow{2}{*}{1008}              & 1177                           & 214                          & 1.672                                              & F, O, Y      &  --   \\
                                      &                      &                        &                                    & 1391                           & 214                          & 1.672                                             & O, Y       &  --   \\
                                      & 02L$^{\star}$                   & 58765.918                  & 5427                               & 1283                           & 642                          & 0.836                                           & D E F G H I J L N O R T Y ab         & \textbf{ac}  \\
                                      & 03L$^{\star}$                   & 58775.076                  & 9000                               & 1283                           & 642                          & 0.836                                           & C D E F G H I J L O R T W Y Z ab         &  --   \\
                                      & 04L$^{\star}$                   & 58792.886                   & 5400                               & 1283                           & 642                          & 0.836                                          & D E F G H I J M O Q R S T W Y Z aa ab          &   --  \\
                                      & 05L$^{\star}$                   & 58831.711                  & 3600                               & 1283                           & 642                          & 0.836                                         & D E F G H I J N O R S T W Y ab          & \textbf{ ad}    \\
                                      & 06L$^{\star}$                   & 58836.783                  & 3600                               & 1283                           & 642                          & 0.836                                          & D F H I L N O R S T W ab          &   --  \\
                                      & 07L$^{\star}$                   & 58849.349                  & 3600                               & 1283                           & 642                          & 0.836                                         & C D F G I J M N O Q R S T ab           &  --   \\
                                      & 08L$^{\star}$                   & 58875.908                  & 3600                               & 1283                           & 642                          & 0.836                                          & C D E F G H I J O R T U Y ab          &  --   \\
                                      & 09U\phantom{*}                   &  58928.738                 & 5400                               & 816                           & 544                          & 0.531                                         & C D E F G H I J L N O Q R S T U W Y Z aa ab         &  --   \\
\hline
M62                                   & \multirow{2}{*}{01L$^{\star}$}                  &  \multirow{2}{*}{58602.820}                  & 4538                               & 1122                           & 321                          & 0.836                                      & A B C D E F              & \textbf{G}    \\
                                      &                   &                   & 4538                               & 1444                           & 321                          & 0.836                                              & A B C D E F      & G    \\
                                      & 02L$^{\star}$                   & 58769.438                  & 9000                               & 1284                           & 642                          & 0.836                                                & A B C D E F    & G    \\
                                      & 03L$^{\star}$                   & 58802.451                  & 9000                               & 1284                           & 642                          & 0.836                                                & A B C D E F    & G    \\
                                      & 04L$^{\star}$                   & 58818.426                  & 9000                               & 1284                           & 642                          & 0.836                                                & A B C D E F    & G    \\
                                      & 05L-1of2$^{\star}$                  & 58844.331                  & 1450              & 1284                           & 642                          & 0.836                                                & A B C D E F    & G    \\
                                      & 05L-2of2$^{\star}$                   & 58844.353                  & 7800               & 1284                           & 642                          & 0.836                                                & A B C D E F    & G    \\

                                      & 06L-orb1$^{\star}$                 & 58895.017                  & 7840                               & 1284                           & 856                          & 0.836                                                & A B C D E F    & G    \\
                                      & 07L-orb2$^{\star}$                 & 58900.996                  & 12600                              & 1284                           & 856                          & 0.836                                                & A B C D E F    & G    \\
                                      & 08L-orb3$^{\star}$                 & 58901.998                  & 4800                               & 1284                           & 856                          & 0.836                                                & A B C D E F    & G    \\
                                      & 09L-orb4$^{\star}$                 & 58902.056                  & 7400                               & 1284                           & 856                          & 0.836                                                & A B C D E F    & G    \\
                                      & 10L-orb5$^{\star}$                 & 58904.023                  & 12600                              & 1284                           & 856                          & 0.836                                                & A B C D E F    & G    \\
                                      & 11L                 & 59074.786                 & 12600                              & 1284                           & 856                          & 0.836                                                & A B C D E F    & G    \\
\hline
NGC 6397 & 01L$^{\star}$                   & 58858.149                  & 9000                               & 1283                           & 642                          & 0.836                                 & A                   &   --  \\
\hline
Ter 5 & 01L$^{\star}$                   & 58630.813                  & 9000                               & 1284                           & 642                          & 0.836                   & All but P, S, ad, ah, aj, al                                &    \textbf{an} \\
                                      & 02L-orb1$^{\star}$                 & 58905.048                  & 12600                              & 1284                           & 856                          & 0.836                                              &   All but J, P, ad, ah, al   &   an  \\
                                      & 03L-orb2$^{\star}$                 & 58906.017                  & 12600                              & 1284                           & 856                          & 0.836                                              & All but P, ad, ah, aj, al     & an    \\
                                      & 04L-orb3$^{\star}$                 & 58907.022                  & 12600                              & 1284                           & 856                          & 0.836                                              & All but A, P, ad, ah, al     &  an   \\
                                      & 05L-orb4$^{\star}$                 & 58907.326                  & 12600                              & 1284                           & 856                          & 0.836                                              & All but P, ad, ah, al     &  an   \\

\hline
NGC 6522 & 01L$^{\star}$                   & 58704.596                  & 9000                               & 1284                           & 642                          & 0.836                                              & A C      &   \textbf{D}  \\
                                      & 02L$^{\star}$                   & 58772.653                  & 9000                               & 1284                           & 642                          & 0.836                                              & A C      &  D   \\
                                     & 03U                & 59046.701                  & 2400                    & 816                           & 544                          & 0.531                                                & A   &  D   \\

\hline
NGC 6624 & 01L$^{\star}$                   & 58736.715                  & 9000                               & 1284                           & 642                          & 0.836                              & A B C D F                      &   \textbf{G}, H  \\
                                      & 02L$^{\star}$                   & 58771.685                  & 17400                              & 1284                           & 642                          & 0.836                                                & A B C D E F    &  G  \\
                                      & 03L$^{\star}$                   & 58796.566                  & 9000                               & 1284                           & 642                          & 0.836                                                & A B C D E F    &  G, H   \\
                                      & 04L$^{\star}$                   & 58870.134                  & 9000                               & 1284                           & 642                          & 0.836                                                & A B C D E F    & G,  \textbf{H}  \\
                                      & 05L$^{\star}$                   & 58878.113                  & 9000                               & 1284                           & 642                          & 0.836                                                & A B C D E F  &  G, H   \\
                                      & 06L-orb1$^{\star}$                 & 58909.030                  & 7200                               & 1284                           & 856                          & 0.836                                                & A B C D E F    & G, H    \\
                                      & 07L-orb2$^{\star}$                 & 58909.406                  & 7200                               & 1284                           & 856                          & 0.836                                                & A B C D E F    & G, H    \\
                                      & 08L-orb3$^{\star}$                 & 58910.023                  & 7200                               & 1284                          & 856                          & 0.836                                                 & A B C D E F    & G, H    \\
                                      & 09L-orb4$^{\star}$                 & 58912.017                  & 7200                               & 1284                           & 856                          & 0.836                                                & A B C D E F    &  G   \\
                                     & 10L-trapum                & 58990.002                  & 14400                    & 1284                           & 856                          & 0.209                                                & A B C D E F  &  G, H   \\
                                     & 11U                & 59046.731                  & 2400                    & 816                           & 544                          & 0.531                                                & A B C D F   &  G   \\
                                     & 12L-trapum                &  59072.807                 & 14400                    & 1284                           & 856                          & 0.209                                                & A B C D E F   & G, H     \\
                                     
\hline
M28      & 01L$^{\star}$                   & 58683.852                  & 9000                               & 1284                           & 642                          & 0.836                                              & A B C D E G H I J K L       & --    \\
                                      & 02L$^{\star}$                   & 58883.106                  & 9000                               & 1284                           & 642                          & 0.836                 & A B C D E G H J K L         & --    \\
\hline
NGC 6752 & 01L-1of3$^{\star}$  & 58666.786 & 3600                               & 1283                           & 642                          & 0.836                                             & B C D E         &  F   \\
                                      & 01L-2of3$^{\star}$                     & 58666.828                       & 3600                               & 1283                           & 642                          & 0.836                                              & B D E        &  --   \\
                                      & 01L-3of3$^{\star}$                     &  58666.870                      & 1800                               & 1283                           & 642                          & 0.836                                             & B D E       & --    \\
                                      & 02L$^{\star}$                   & 58850.637                  & 9000                               & 1283                           & 642                          & 0.836                                             & B D &  \textbf{F}   \\
                                      
\hline
M30      & 01L$^{\star}$                   & 58750.786                  & 11100                              & 1283                           & 642                          & 0.836                                               & A     & --   \\
                                      & 02L$^{\star}$                   & 58844.526                  & 9000                               & 1283                           & 642                          & 0.836                                             & A       & --    \\
\hline
\end{tabular}\\
\end{table*}

\subsection{Setup and observing strategy}
The nine selected GCs were observed from 2019 March to 2020 August. The full list of observations used for this work is reported in Table \ref{tab:list_observations}. For almost all the observations, we used the L-band (856--1712 MHz) receivers and up to 42 of the 44 antennas that are located within a radius of 1 km from the nominal phase center of the array. We will refer to this subset of the MeerKAT antennas as the ``1-km core'' through the rest of the paper. The choice of using only the 1-km core antennas was dictated by the need to have a large enough beam on the sky: at the central frequency $\fc = 1284$~MHz of the L band, each semi-axis of the single tied-array beam has a minimum size of $\sim 0.5$~arcmin, enough to cover the central regions of most of the observed GCs.
With the exception of some data collected at early stages of the experiment, the data were acquired in PSRFITS search-mode format \citep{Hotan+2004} with the MeerTime PTUSE backend. Initially, we used a nominal bandwidth of 642 MHz, divided into 768 channels. From 2020 February the nominal bandwidth covered the full 856 MHz provided by the L-band receivers, and was divided into 1024 channels, hence retaining the same channel width of 0.836 MHz. The band was coherently de-dispersed using the median DM of the known pulsars of the cluster and sampled every 9.57 $\mu$s, retaining all four Stokes parameters.

We observed each cluster at least once with an integration time of 2.5 hours. After that, depending on whether one or more potential candidate new pulsars were found, the same cluster was re-observed once or a few more times, with the same setup and integration time whenever possible.
The only exception to this strategy was applied to 47 Tuc. In the latter cluster, the observed flux densities of the pulsars are highly dominated by scintillation and show large variations from day to day. Hence, for 47 Tuc, we opted for several, shorter ($\sim1$~h) observations, regardless of the candidates that they produced each time. This was done to increase the likelihood of discovering pulsars whose signals were by chance boosted by scintillation above our sensitivity limit.

For a few clusters, we also performed additional observations aimed at following up some of the discoveries. These include ``orbital campaigns'' (see Section \ref{sec:orbital_determination}), needed to determine the orbital parameters of binary systems, as well as observations made with the UHF receivers (550--1050 MHz), needed to assess the best observing band for long-term follow-up timing. Motivated by the discovery of a highly eccentric binary MSP in NGC 6624 (pulsar G, see Section \ref{sec:discoveries_NGC6624}), we also re-observed the latter cluster on two occasions making use of the tiling capabilities of the TRAPUM backend, APSUSE. The main purpose of these observations was the localization of several pulsars (including new discoveries) that were lacking a coherent timing solution, and hence, a known position, in NGC~6624. The TRAPUM backend was used in combination with the full MeerKAT array. This is made by up to 64 antennas (but 60 were available during the observations) that have a maximum baseline of 7.7 km. This translates into a beam diameter at half-power of $\sim 12$~arcsec at 1.3 GHz. For the NGC~6624 observations, we synthesized 288 coherent beams with a >70\% overlap between them. The resulting beam tiling covered an area of $\sim 2$~arcmin in radius around the cluster's nominal center. The data were recorded in search mode, with a 76-$\mu$s time resolution, 856 MHz of bandwidth centered at 1284 MHz divided into 4096 frequency channels with no coherent de-dispersion, and integrated for 4 hours. 
We point out that, for this work, we focused on the search for new pulsars only in the MeerTime L-band observations performed until 2020 July, including those performed for the orbital campaigns. 
The aforementioned TRAPUM L-band and MeerTime UHF-band observations are only reported here for their timing implications, whereas the search for new pulsars in those data sets will be discussed elsewhere.

\subsection{Sensitivity}
The observing setup of the MeerTime L-band observations described above translates into a nominal sensitivity that can be estimated with the radiometer equation \citep{Dewey+1985}. The latter returns the mean flux density, $S_{\rm mean}$, that a pulsar must have to be detected with a given signal-to-noise ratio (S/N) as a function of the observing parameters, as:
\begin{equation}
\label{eq:radiometer_equation}
S_{\rm mean} = \frac{ {\rm S/N} \,\,  \beta  \,\, T_{\rm sys}} { G \sqrt{n_{\rm pol}  \  \BWeff ~ \Delta t_{\rm obs}}}  \,\, \sqrt{\frac{\zeta}{1 - \zeta}} \,.
\end{equation}

Here $G$ is the telescope gain, which, when using the 42 antennas of the typical observations made with the MeerKAT 1-km core, is $1.84$~K\,Jy$^{-1}$.  $T_{\rm sys} = 26$~K is the total system temperature for L-band observations, and is the sum of several contributions: the receiver temperature $T_{\rm rec}$ (18~K), the cold sky temperature $T_{\rm sky}$ ($\sim3.5$~K at 1.3 GHz), and the atmosphere plus the ground spillover temperature $T_{\rm atm+spill}$ ($\sim4.5$~K at 45 deg elevation\footnote{ \url{https://skaafrica.atlassian.net/rest/servicedesk/knowledgebase/latest/articles/view/277315585}}). $n_{\rm pol}=2$ is the number of orthogonal polarizations summed in the searched signal; $\beta$ is a factor that accounts for various sensitivity losses due to signal processing and digitization, and we assume it to be 1.1; $\BWeff \simeq 650$~MHz is the typical effective observing bandwidth, after the removal of frequency channels polluted by radio frequency interference (RFI); $\Delta t_{\rm obs}$ is the integration time, which we take to be 2.5 h, as for our typical survey observations; $\zeta$ is the observed pulse duty cycle, resulting from the convolution of the intrinsic pulse width of the pulsar with both the effects of the interstellar medium and the signal acquisition system. 
In order to give a value of the $S_{\rm mean}$ for this GC census survey, and to allow a comparison with other pulsar search experiments, we adopt a minimum ${\rm S/N}=10,$ and a $\zeta=8$~per cent. With the above parameters, we are sensitive to slow and mildly recycled pulsars (with spin periods $P \gtrsim 10$ ms) down to a mean flux density of $S_{\rm mean}\sim 13$~$\mu$Jy at 1.3 GHz, at the beam boresight. For the fastest MSPs ($P \lesssim 10$~ms), the miminum detectable $S_{\rm mean}$ is higher essentially because of scattering broadening, which can become important at the relatively low frequencies of our L-band observations. Other effects, such as intrachannel dispersive smearing, are negligible thanks to coherent de-dispersion (see Section 2.3 of \citealt{Hessels+2007} for a detailed discussion of all the effects that can impact $\zeta$). Figure~\ref{fig:sensitivity} compares the sensitivity curve of this initial MeerKAT GC census survey, against those of other major GC pulsar surveys, conducted at Parkes \citep{Possenti+2005}, Arecibo \citep{Hessels+2007} and the GBT \citep{Ransom+2004,DeCesar+2015}, with either their L-band ($\fc \sim 1.4$~GHz) or S-band ($\fc \sim 2$~GHz) receivers. For clusters located in the far South ($\delta \lesssim -45^\circ$), we have $\gtrsim 5$ better sensitivity than previous surveys carried out at Parkes. For clusters with higher declinations ($-45 \lesssim \delta \lesssim 0^\circ$) our sensitivity is a factor of $\sim 2$ better than the GBT L-band survey. This can be even higher for those clusters only visible with very low elevations by the GBT (e.g. NGC 6522, NGC 6624, M62), for which atmospheric opacity is significant and increases the total $T_{\rm sys}$ by several K.


\section{Analysis}
\label{sec:analysis}

\subsection{Data reduction}
As a first step in our analysis, we used the \texttt{psrfits\_subband} routine, part of the \texttt{PSRFITS\_UTILS}\footnote{\url{https://github.com/scottransom/psrfits\_utils}} package, to downsample each observation in both frequency and time, and to sum the two polarizations, thus keeping total intensity only. While doing so, the bandpass was flattened using the proper scale and offset values, which were recorded while acquiring the data. For all the observations we retained a time resolution of 76 \us. The number of frequency channels was chosen to be such that, in the worst case scenario of a new pulsar having a DM that is 5\% off the median DM of all the other pulsars known in the cluster, the intrachannel smearing would anyway be less than 76 \us, preventing dispersive smearing from dominating the loss of sensitivity due to pulse broadening.
This reduced the size of the data to search from $\sim 1100$~GB/h for the native-resolution files, to only $\sim 6-18$~GB/h, depending on the cluster.

\subsection{Pulsar searching}
The searching analysis was carried out using \PULSARMINER\footnote{\url{https://github.com/alex88ridolfi/PULSAR\_MINER}}~v1.1, an automated pipeline built upon the \PRESTO\footnote{\url{https://www.cv.nrao.edu/~sransom/presto/}} v2.1 pulsar searching package \citep{Ransom+2002}. All the routines mentioned in this Section are part of \PRESTO, unless otherwise stated.

First, a list of bad frequency channels and time intervals in the observations were flagged as contaminated by RFI and excluded from the rest of the analysis. This was done by generating a mask with the \texttt{rfifind} routine. In particular, some of the frequency channels that are notoriously polluted by strong RFI were marked manually. Additional bad channels and time intervals were automatically flagged by \texttt{rfifind}, based on statistical analyses. 

Ignoring the masked channels, a 0-DM time series was created using \texttt{prepdata}. A Fourier transform of the latter led to the detection of prominent periodic RFI (the so called  ``birdies'') still present in the data. The frequencies of such signals were stored in a ``zaplist'' file, to be later removed from the Fourier spectra where new pulsars are searched. Similarly, for all the previously known isolated pulsars in the cluster, the barycentric spin frequencies and their harmonics (up to eight) were calculated at the epoch of the observation and listed in the same zapfile, to later be also removed from the Fourier spectra. This is especially useful to prevent the re-detection of very bright isolated pulsars, which are likely to produce a large number of candidates, so as to save computational time that would otherwise be needed for their folding.

Using the \texttt{DDplan.py} routine, an optimal de-dispersion scheme was generated. This is automatically computed on the basis of the DM range to search for the particular cluster, the DM value at which the data were coherently de-dispersed, as well as on the other observing parameters (see e.g. \citealt{Lorimer_Kramer2004}). The range of DMs searched for each cluster was chosen to be between $\pm5$~\% of the median DM of the known pulsars of that cluster. The mask and the de-dispersion scheme were fed into \texttt{prepsubband} to generate a number of (mostly) RFI-free time series, de-dispersed at the different DM trial values within the chosen range, and referred to the Solar System barycenter.
The barycentered time series were then Fourier transformed with \texttt{realfft} and the resulting power spectra de-reddened using \texttt{rednoise}. 

The periodicity search was performed using a matched filtering algorithm in the Fourier domain, implemented in \texttt{accelsearch}. 
While an isolated pulsar has its power concentrated in just the Fourier bin associated with its spin frequency and its harmonics, this is not the case for a binary pulsar. In fact, for an observation of length $\Delta t_{\rm obs}$, a pulsar with a spin period $P$ that is undergoing a line-of-sight acceleration $a_l$ due to an orbital motion, will see the power associated with its spin frequency shift in the Fourier domain by a number $z = \Delta t_{\rm obs}^2 a_l / cP$ of Fourier bins, where $c$ is the speed of light. \texttt{accelsearch} is capable of recovering such Doppler-shifted periodic signals of binary pulsars, by also considering the Fourier powers spread over up to $\zmax$ bins around a given frequency, regarded as the fundamental harmonic. 
For our searches, we used a value of $\zmax = 0$ for targeting new isolated pulsars, and a $\zmax = 200$ for possible new binary pulsars. In both cases, the powers of the first eight harmonics were summed.
It is important to note that the main assumption of the algorithm is that the acceleration along the line of sight undergone by the pulsar during the observation be constant. This assumption is not valid when $\Delta t_{\rm obs} \gtrsim 0.1 \Pb$, where $\Pb$ is the binary orbital period \citep{Ransom+2003}. For this reason, we  also performed a so-called ``segmented search'' \citep{Johnston_Kulkarni1991}, where the observation is split into shorter sections, each of which is searched individually. This allows for the detections of pulsars in tighter binaries, at the cost of a reduced sensitivity caused by the shorter integration time. Hence, besides searching the full-length observation (which gives the maximum sensitivity to isolated pulsars and wide binaries), we split each observation into sections of 60, 30, 15 and 5 minutes (with no overlaps) and searched each of them individually. This strategy made our search potentially sensitive to bright pulsars in extremely compact binary systems with orbital periods as short as $\sim50$~minutes.

\subsection{Candidate sifting and folding}
For each section, all the candidates produced by the acceleration search were grouped according to their harmonic relations and sifted using standard criteria (e.g. the same candidate should be significant in at least three adjacent DM trial values) and kept candidates with a Fourier significance of $4\sigma$ or more (see \citealt{Ransom+2002} for details). This resulted in a number of candidates in the range of 200--3000 for each observation, depending on its length, the number of bright, previously known binary pulsars (whose periodicities were not removed from the Fourier spectra) in that cluster and the residual RFI contamination.
All the candidates were automatically folded with \texttt{prepfold}, using their nominal period, DM and possible acceleration, and allowing the software to optimize those parameters and maximize the $\chi^2$ of the folded profile \citep{Leahy+1983}. The folds produced diagnostic plots containing, among other things, the integrated pulse profile, and the signal intensity as a function of time and frequency. All the plots were inspected visually. The ones with a high $\chi^2$, persistent and broadband signals were marked as ``good'' candidates, and further investigated.

\subsection{Confirmation and characterization}
In order to be proclaimed as a newly discovered pulsar, a candidate (i.e. a $P$--DM pair) was required to be identified in at least two separate observing epochs of the same cluster. For some of the low-DM clusters, where the pulsars scintillate, we also used archival data taken at other radio telescopes to confirm the new discoveries. In that case, we folded the archival data (and then optimized the result) using the $P$--DM values of the new pulsar; when no detection resulted from direct folding (as is often the case for binary pulsars) we used \texttt{accelsearch} to search for the signal within a narrow range of $P$ and DM around their nominal values.

\subsubsection{Orbital determination for binary pulsars}
\label{sec:orbital_determination}

If the observed spin period, $\Pobs$, and possibly the associated observed spin period derivative, $\Pdotobs$, of a newly found pulsar appeared different from epoch to epoch, we considered that as a sign of binarity of the pulsar. To estimate the orbital parameters, we started following the method of \citet{Freire+2001}. First, we measured the barycentric $\Pobs$ and $\Pdotobs$ from all the available detections of the new pulsar. If the detection significance of a particular observation was high, we split that observation into sections, so as to have more $\Pobs$ and $\Pdotobs$ measurements. We then converted each $\Pdotobs$ into the corresponding line-of-sight acceleration, $a_l$, via the relation $a_l = c (\Pdot / P)_{\rm obs}$, and plotted it as a function of $\Pobs$. The resulting plot is referred to as ``Period-Acceleration'' diagram, where the $(\Pobs, a_l)$ points will fall on a closed curve, whose position and shape are related to the pulsar's intrinsic spin period, $\Pint$, and to the orbital parameters. In the case of a circular orbit, the points follow an ellipse, which can be fitted to obtain estimates of the orbital period, $\Pb$, and of the semi-major axis of the pulsar orbit, projected along the line of sight, $\xp$. If the orbit is eccentric, the curve deviates from a perfect ellipse, and its more complex shape embodies information on the eccentricity, $e$, the longitude of periastron of the pulsar orbit, $\omega$, as well as on $\Pb$ and $\xp$.

For the new binary pulsars that were not affected by scintillation (i.e. their flux density appeared stable in each detection), we performed a few additional observations that were aimed at refining their orbital parameters. We regard these observations as ``orbital campaigns'' (and label them with ``-orbX'' in Table \ref{tab:list_observations}).
These orbital campaigns consisted of 4-5 observations, performed within a time range of a few days, with long enough integration times to guarantee a firm detection of the pulsar in each of them. 
With these detections, we could fit the observed spin period as a function of time, $\Pobs(t)$. We did so using some of the python scripts included in \PRESTO, namely \texttt{fit\_circular\_orbit.py} for the circular binaries, and \texttt{fit\_orb.py} for the eccentric ones. Using the results of Period-Acceleration diagram as starting estimates, the $\Pobs(t)$ fit returned much improved orbital parameters, which could then be further refined by fitting the pulse times-of-arrival (ToAs, see next Section). 

For two new circular binaries that were affected by scintillation, this strategy was not viable. For them, we used the ``periodogram'' method (see e.g. \citealt{Ridolfi+2016}) to refine the measurement of $\Pb$. On each observation where the pulsar was detected, we used \SPIDERTWISTER\footnote{\url{https://github.com/alex88ridolfi/SPIDER_TWISTER}} to perform a brute-force search in orbital phase, assuming the initial $\Pb$ and $\xp$ estimated through the Period-Acceleration diagram. The search returned the best times of ascending node, $\Tasc$, closest to the epoch of each detection. Exploiting the fact that each $\Tasc$ pair must accommodate an integer number of orbits, this method allowed us to obtain a much improved estimate of $\Pb$.

\begin{table*}
\caption{List of the newly discovered pulsars and their main characteristics.}
\label{tab:discoveries}
\footnotesize
\centering
\renewcommand{\arraystretch}{1.0}
\vskip 0.1cm
\begin{tabular}{lcclrcccccc}
\hline
\hline
\multicolumn{10}{c}{Summary of Discoveries}\\
\hline
Pulsar   & $P$ & DM         & Type &  $\Pb$   & $\xp$  & $M_c^{\rm min}$ & $e$ & $S_{\rm 1300}$  & Phase-connected   \\
name       & (ms) & (\dmunit) & &  (d)     & (lt-s)  & (\msun) &  & ($\mu$Jy) & solution? &  \\
\hline
47 Tuc ac   &  2.74   & \phantom{0}24.46   & Binary (Black Widow)  &   $\sim 0.18$    & $\sim 0.019$      & $\sim 0.0079$   & 0$^{a}$         & $152^{\rm b}$ & No       \\ 
47 Tuc ad   &  3.74   & \phantom{0}24.41   & Binary (Redback)   &  \phantom{0}0.32           & 0.68      & 0.205  &  0$^{a}$  & $84^{\rm b}$ & No       \\
M62G        &  4.61   & 113.68 & Binary (Pulsar-WD)  & 0.77        & \phantom{0}0.62   & 0.099  & 0.0009   & 80(17) & Partial   \\
Ter 5 an    &  4.80   & 237.74 & Binary (Pulsar-WD) & 9.62       & 12.78   & 0.433   &  0.0066  & 31(2) & Yes\\
NGC 6522D   &  5.53   & 192.73             & Isolated  & --   & --     & --   & --         & 28(9)   & No   \\
NGC 6624G   &  6.09   & \phantom{0}86.21   & Binary (Pulsar-WD/NS)  &  \phantom{0}1.54       & 3.00   & 0.332   & 0.3805   & \phantom{0}47(10) & Yes\\
NGC 6624H   &  5.13   & \phantom{0}86.85   &  Isolated & --    & --     & --  & --          & 31(9)   & No   \\
NGC 6752F   &  8.48   & \phantom{0}33.20   &  Isolated & --   & --     & --   & --         & $55(8)^{\rm b}$   & Yes   \\
\hline
\multicolumn{10}{l}{$^{\rm a}$ Manually set to zero, not result of fit. $^{\rm b}$ Likely biased by scintillation.}
\end{tabular}
\end{table*}

\subsubsection{Timing}

Using the \DSPSR\footnote{\url{http://dspsr.sourceforge.net}} pulsar software package \citep{vanStraten_Bailes2011} we folded all the MeerKAT observations of the GC where a new pulsar was detected, using the nominal $P$--DM values. Whenever possible, we also used archival data taken by other telescopes. 
We used the \PSRCHIVE\footnote{\url{http://psrchive.sourceforge.net}} package \citep{Hotan+2004,vanStraten+2012} tools to clean the resulting folded archives and to extract topocentric ToAs, using a sensible decimation scheme (i.e. choosing a suitable number of subintegrations and frequency channels) for each individual epoch. We then constructed a starting ephemeris with a basic timing model, which included the nominal host cluster's center position, the nominal pulsar barycentric spin frequency, its DM and, if binary, a Keplerian orbital model, with a theory-independent description of relativistic effects, the ``DD'' model \citep{damour_deruelle_86}. The initial Keplerian parameters are derived as
described in Section \ref{sec:orbital_determination}.
This ephemeris was fed to the \TEMPO\footnote{\url{http://tempo.sourceforge.net}} pulsar timing software, which we used to fit the ToAs for the timing model parameters, as well as for arbitrary phase offsets (so-called "jumps") between epochs. Following the procedure described in \citet{Freire_Ridolfi2018}, we started connecting groups of ToAs, removing the arbitrary jumps whenever possible, and updating the model. For the faintest pulsars, and for those with sparse detections, we did this with the help of \texttt{DRACULA}\footnote{\url{https://github.com/pfreire163/Dracula}}, an automated software for determining the correct rotation counts of pulsars. Depending on the pulsar, the precision of the ToAs and the total time spanned by the dataset, we sometimes included additional parameters to the model, such as spin frequency derivatives, the proper motion and post-Keplerian (PK) effects.
The whole procedure was iterated a few times until either all the ToAs were phase-connected or no further improvement of the timing solution was possible.


\section{Results}
\label{sec:results}

\subsection{Discoveries}
\label{sec:discoveries}
\begin{figure*}
\centering
	\includegraphics[width=\textwidth]{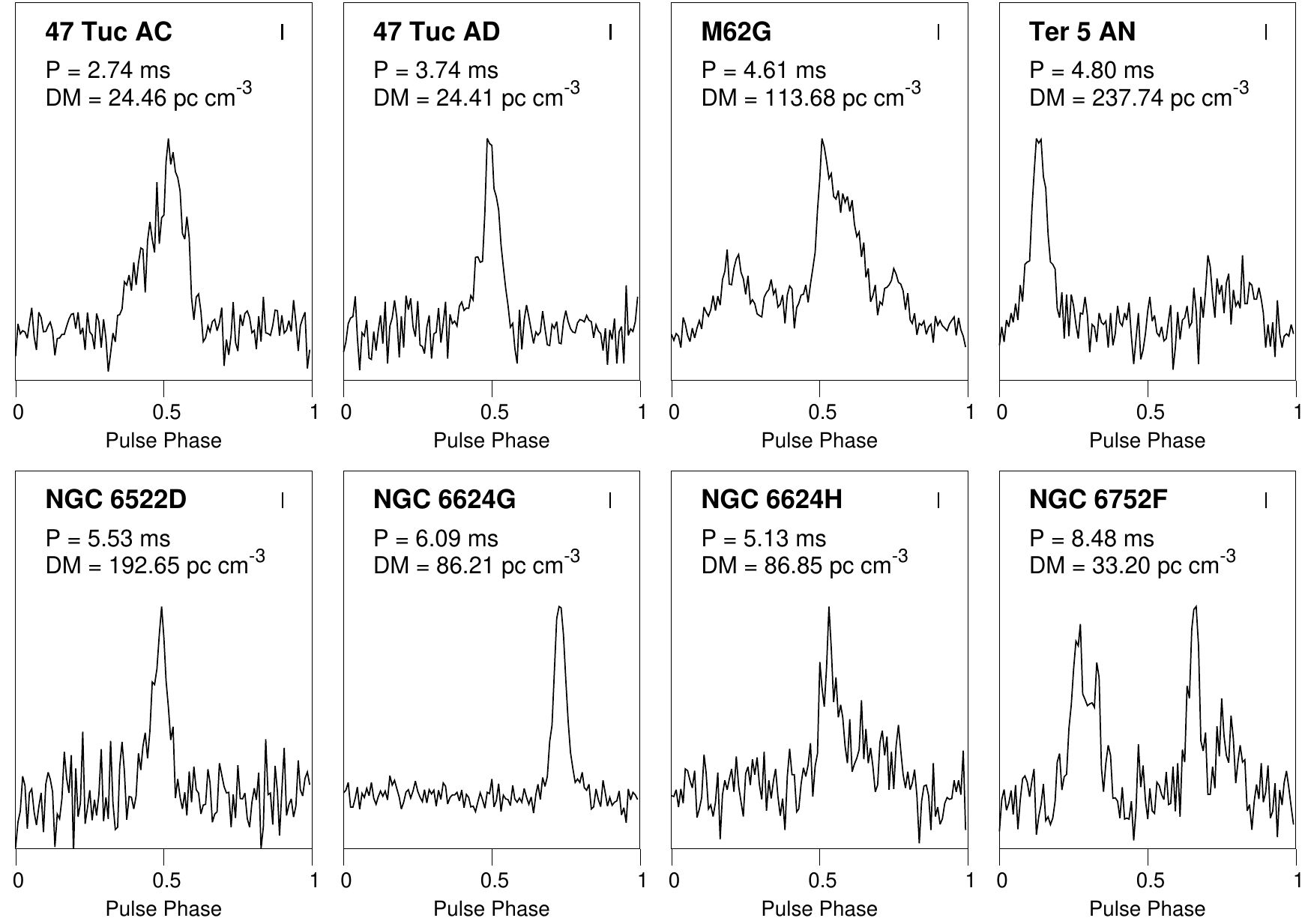}
  	\caption{Integrated pulse profiles of the eight MSPs discovered, with their spin period and DM indicated. The horizontal section of the tiny bars on the top-right of each panel shows the sampling time of the native-resolution (sampling time of 9.57~$\mu$s) MeerTime search-mode data.  }
  	\label{fig:integrated_profiles}
\end{figure*}

The search of the MeerKAT L-band data has resulted in the discovery of eight previously unknown MSPs in six different clusters. Five of the discoveries belong to binary systems, with orbital periods ranging from a few hours to several days. The two most compact binaries also show eclipses.  The spin periods of the new pulsars are all in the very narrow range of $2.74-8.48$~ms. This is not a selection effect, since previously known pulsars with longer periods up to 405 ms were also blindly re-detected by our search pipeline. The main characteristics of the new pulsars, including their estimated mean flux densities, are reported in Table~\ref{tab:discoveries}, whereas their integrated pulse profiles are shown in Figure \ref{fig:integrated_profiles}. In the following, we review each discovery in detail. 

\begin{figure*}
\centering
	\includegraphics[width=0.158\textwidth]{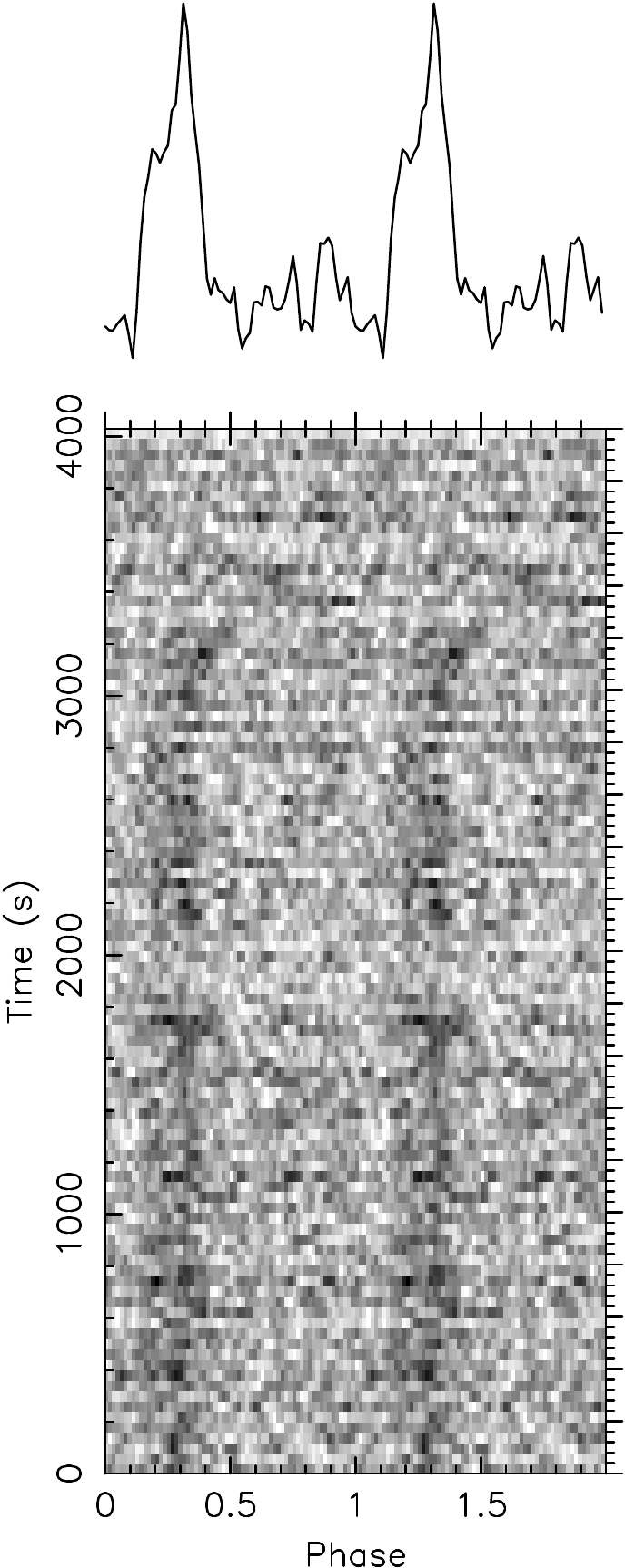}
	\,
	\includegraphics[width=0.158\textwidth]{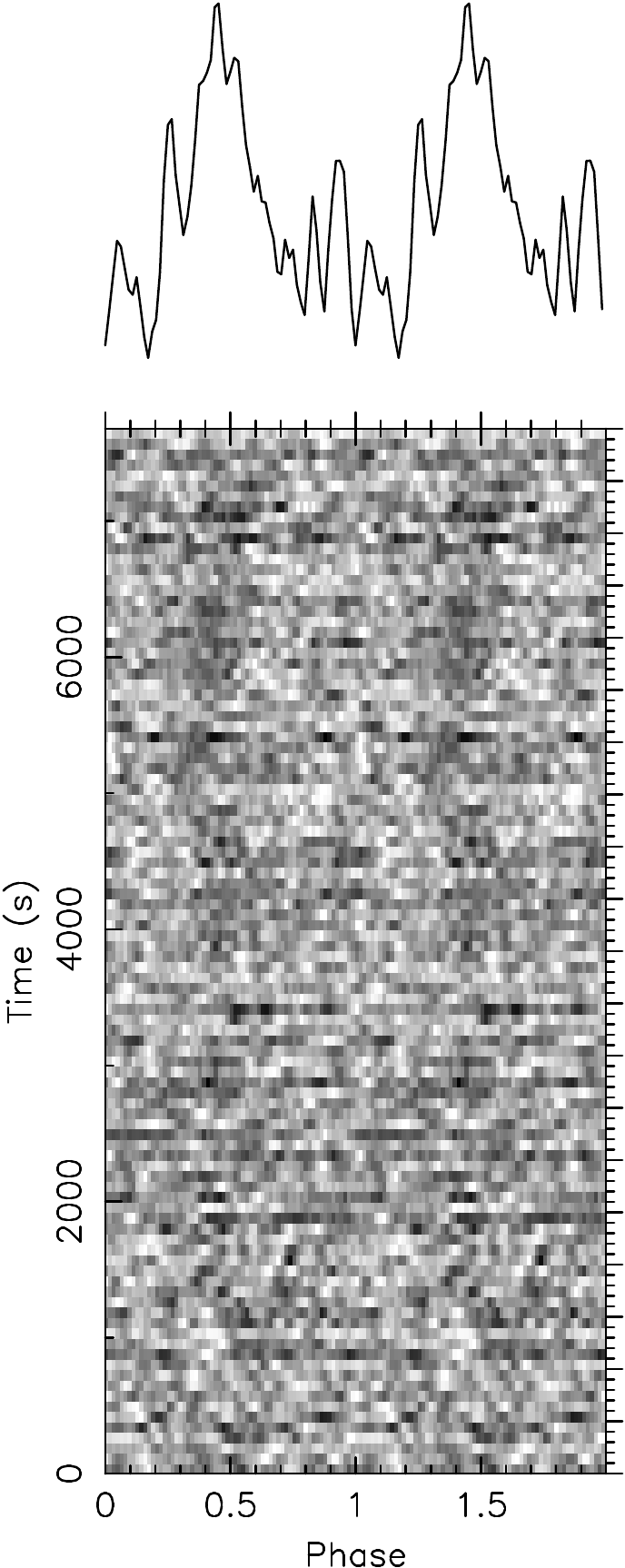}
	\,
	\includegraphics[width=0.158\textwidth]{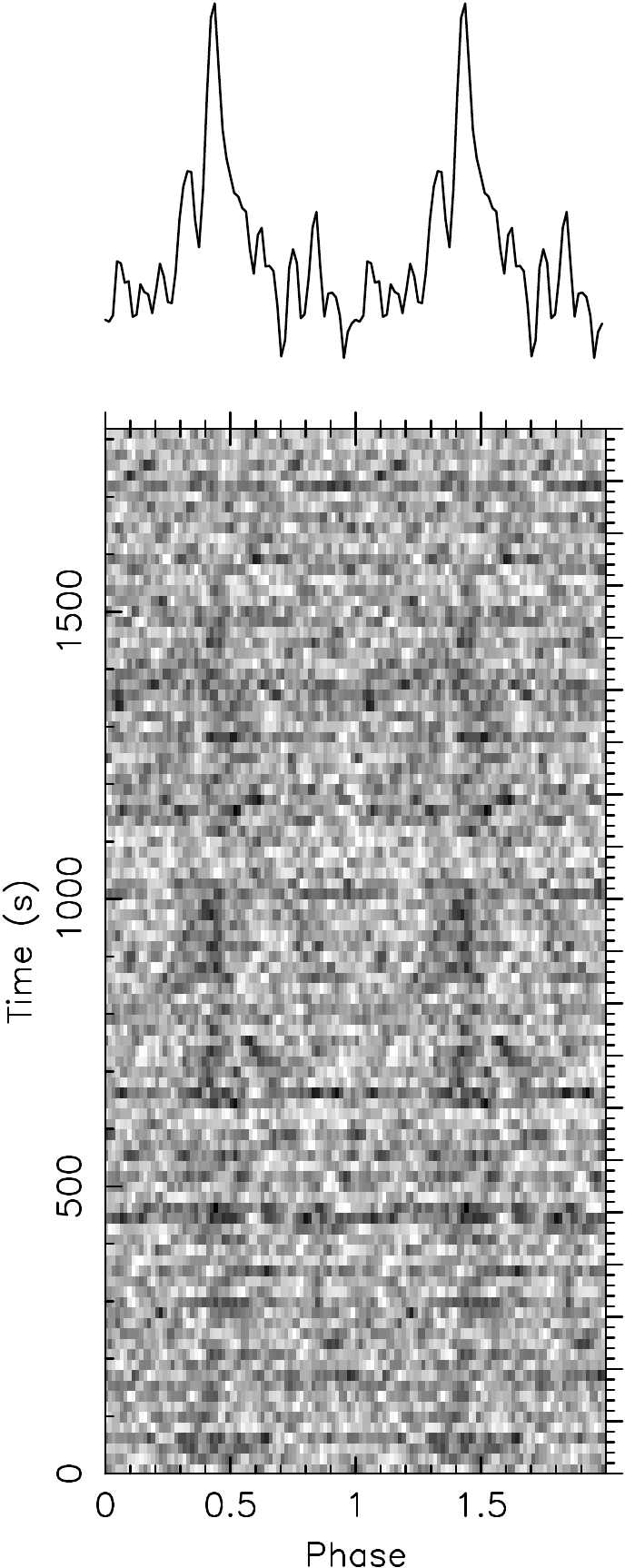}
	\,
    \includegraphics[width=0.158\textwidth]{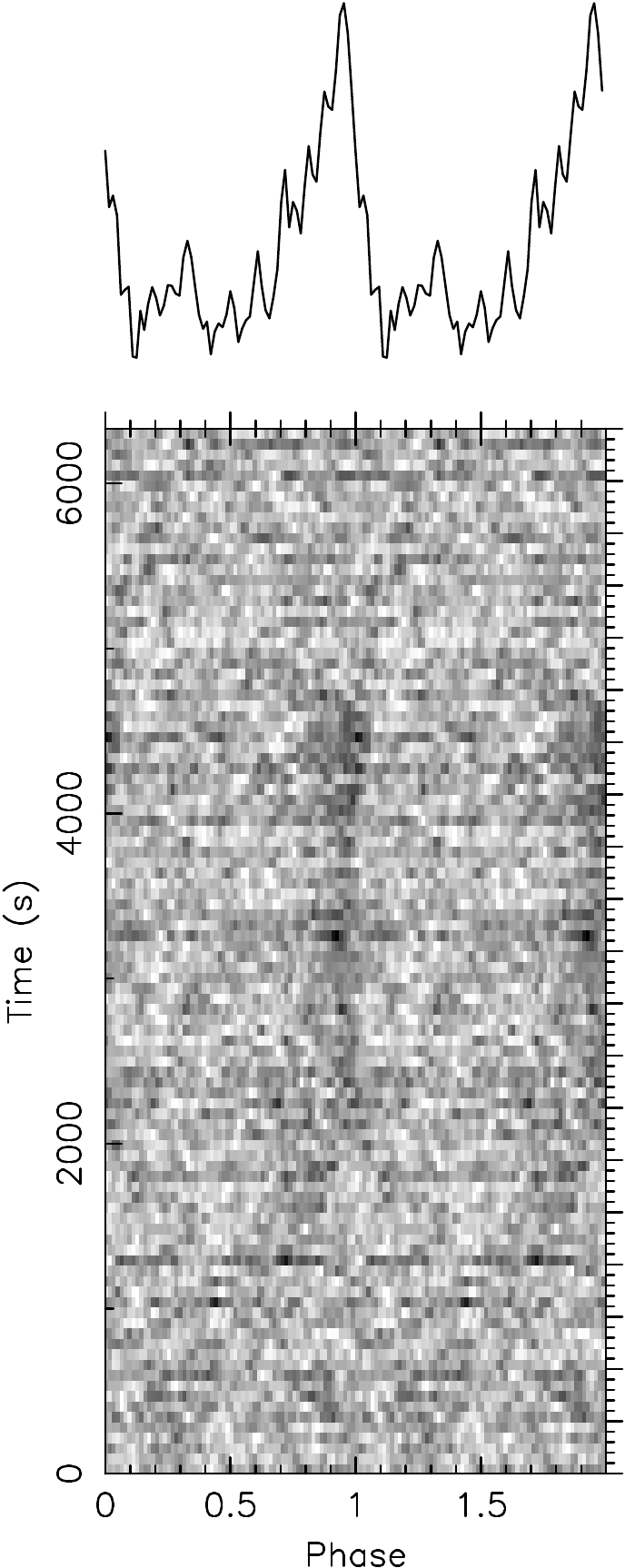}
	\,
	\includegraphics[width=0.158\textwidth]{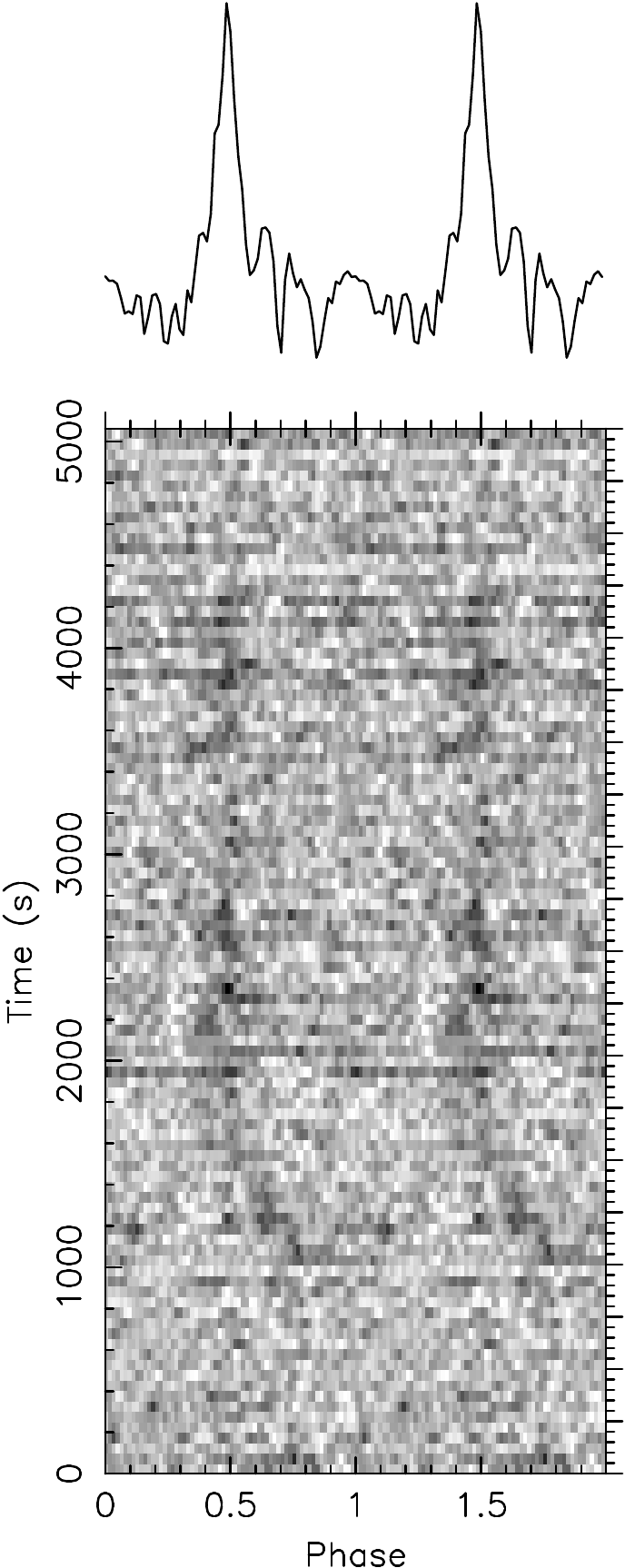}
	\,
	\includegraphics[width=0.158\textwidth]{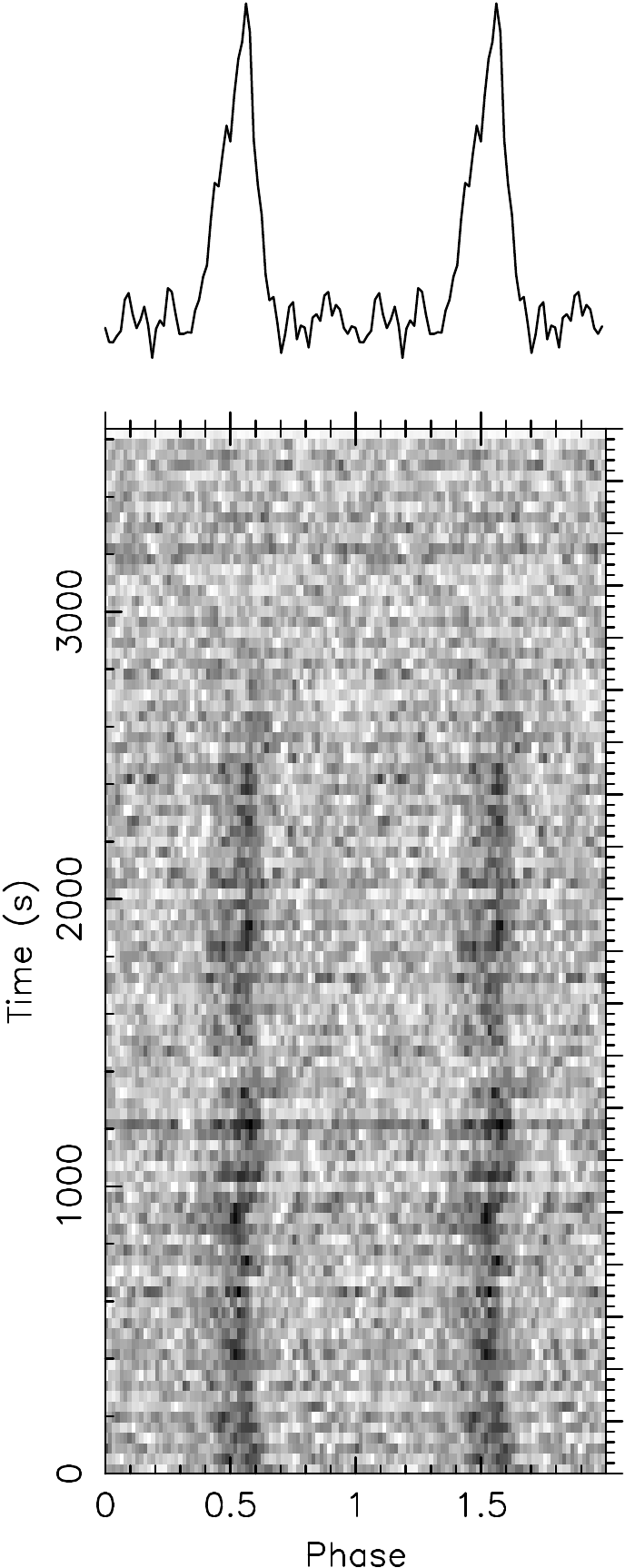}
  	\caption{Intensity as a function of time and spin phase for all the six detections obtained for the new eclipsing black widow pulsar 47~Tuc~ac. From left to right, the first five panels show the detections obtained from Parkes observations made on 2000/05/17, 2000/10/01, 2001/07/27, 2003/10/04 and 2008/07/01, respectively; the rightmost panel shows the MeerKAT discovery observation (id. 02L). The phase drifts shown are likely due to the additional dispersive delay caused by the intra-binary eclipsing material, as well as to some possible residual uncorrected orbital motion due to the poorly determined binary parameters.}
  	\label{fig:47TucAC_folds}
\end{figure*}

\begin{figure}
\centering
    \includegraphics[width=0.44\columnwidth]{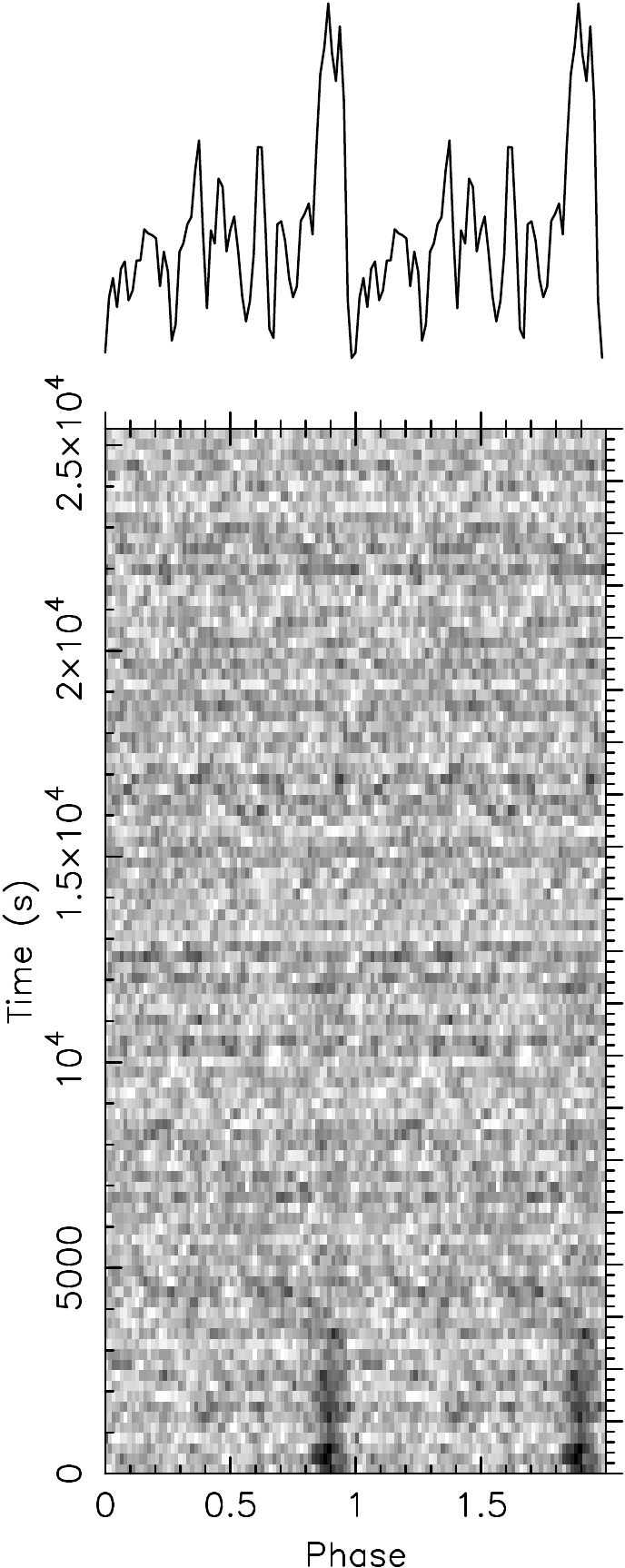}
    \qquad
	\includegraphics[width=0.44\columnwidth]{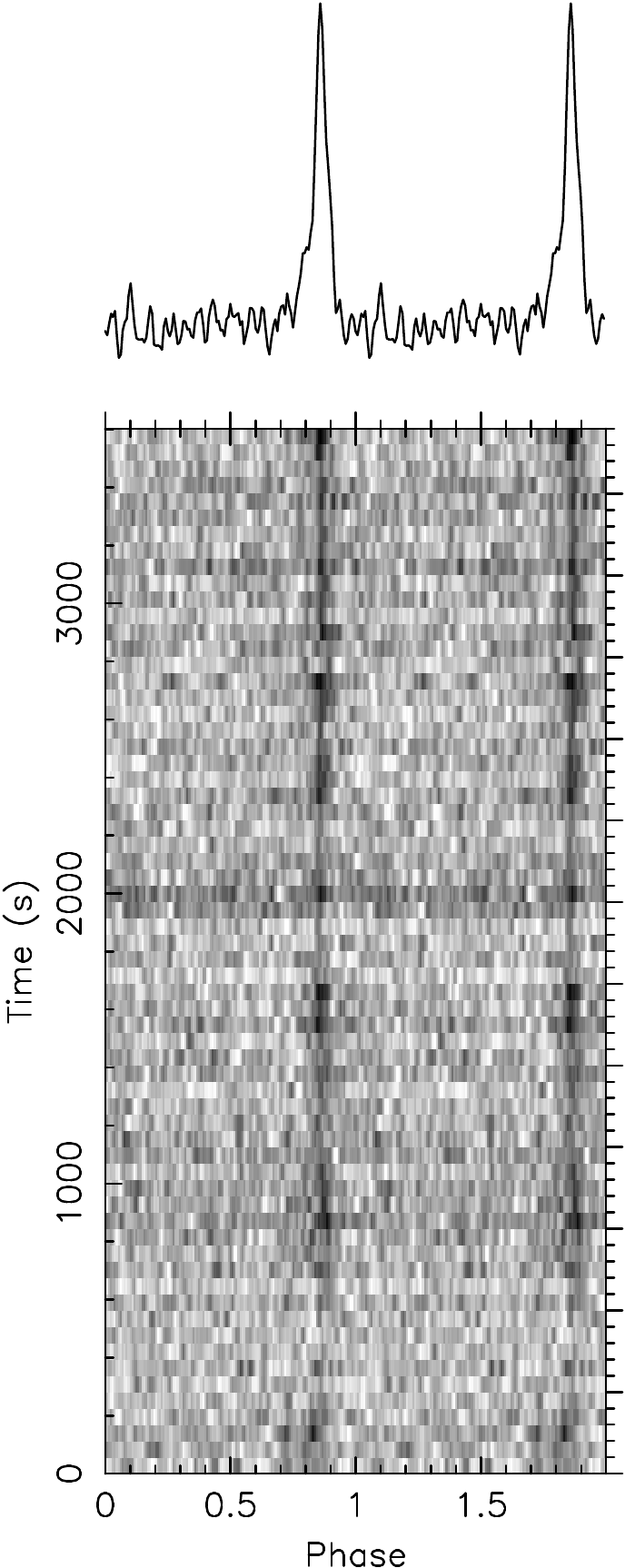}
  	\caption{ Intensity as a function of time and spin phase for the two detections of the new eclipsing redback pulsar 47~Tuc~ad. Left panel: Parkes observation made on 2004/05/27. Right panel: MeerKAT discovery observation (id. 05L).}
  	\label{fig:47TucAD_folds}
\end{figure}

\begin{figure*}
\centering
		\includegraphics[width=0.305\textwidth]{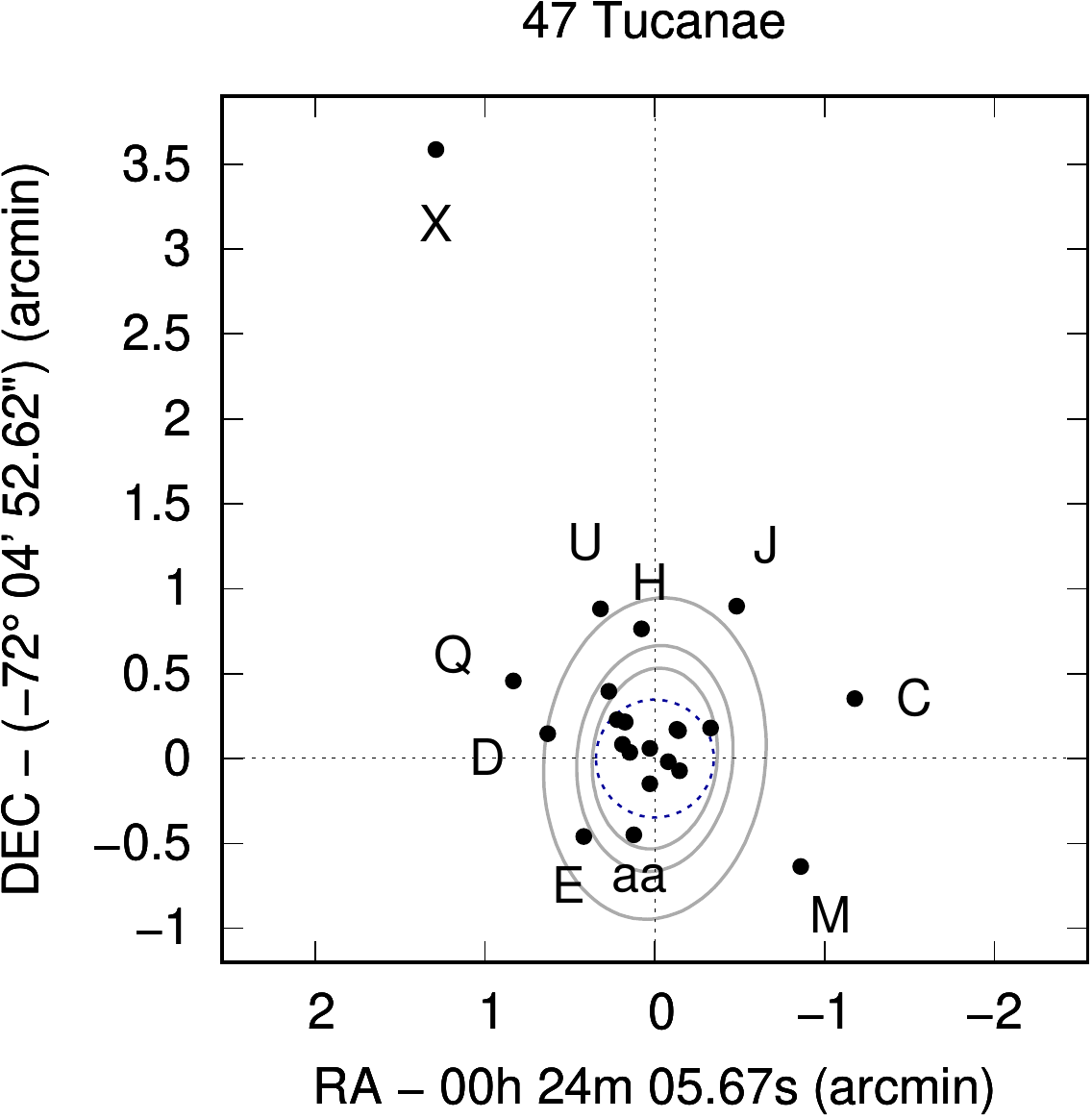}
		\qquad
		\includegraphics[width=0.305\textwidth]{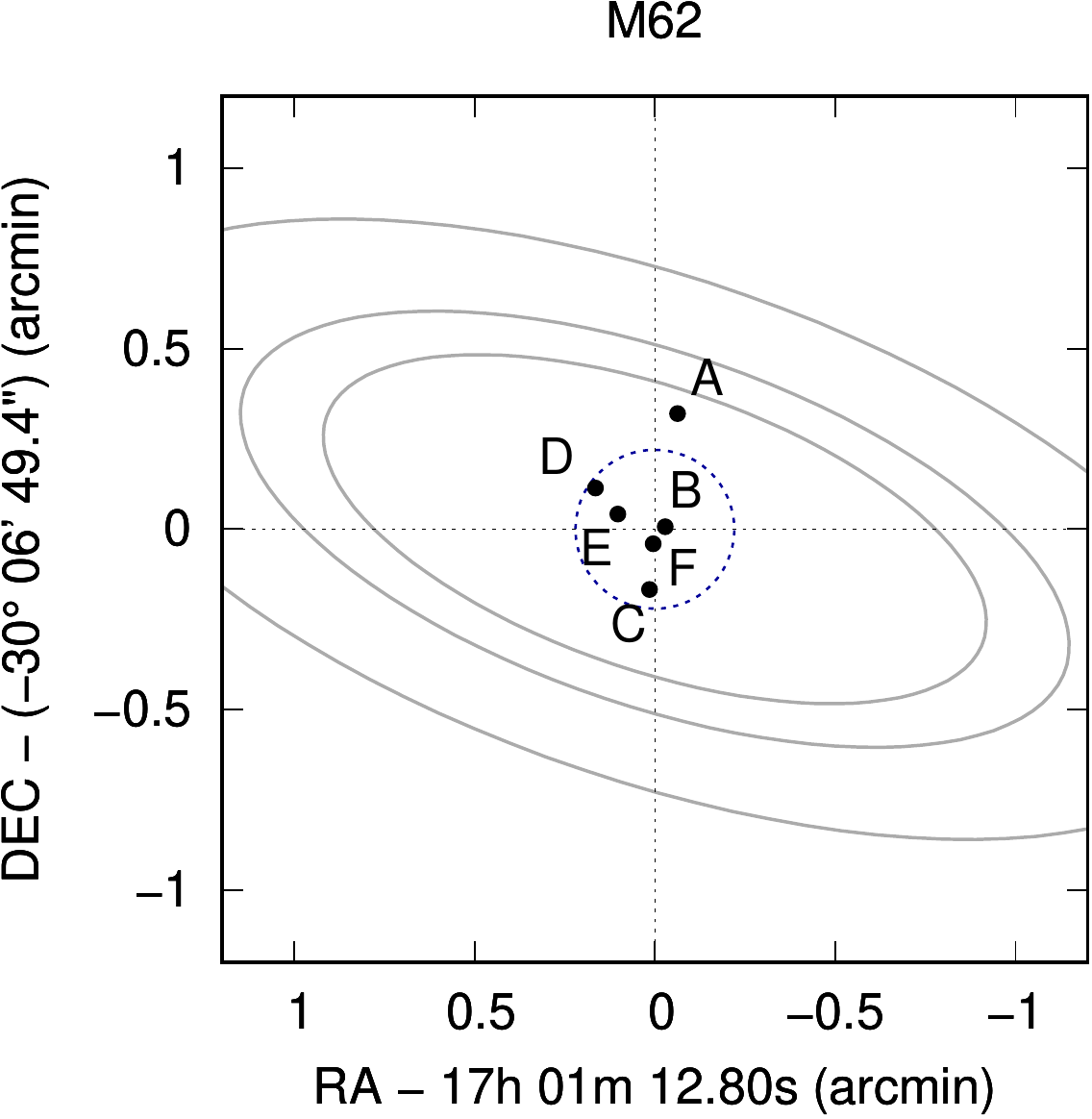}
		\qquad
		\includegraphics[width=0.305\textwidth]{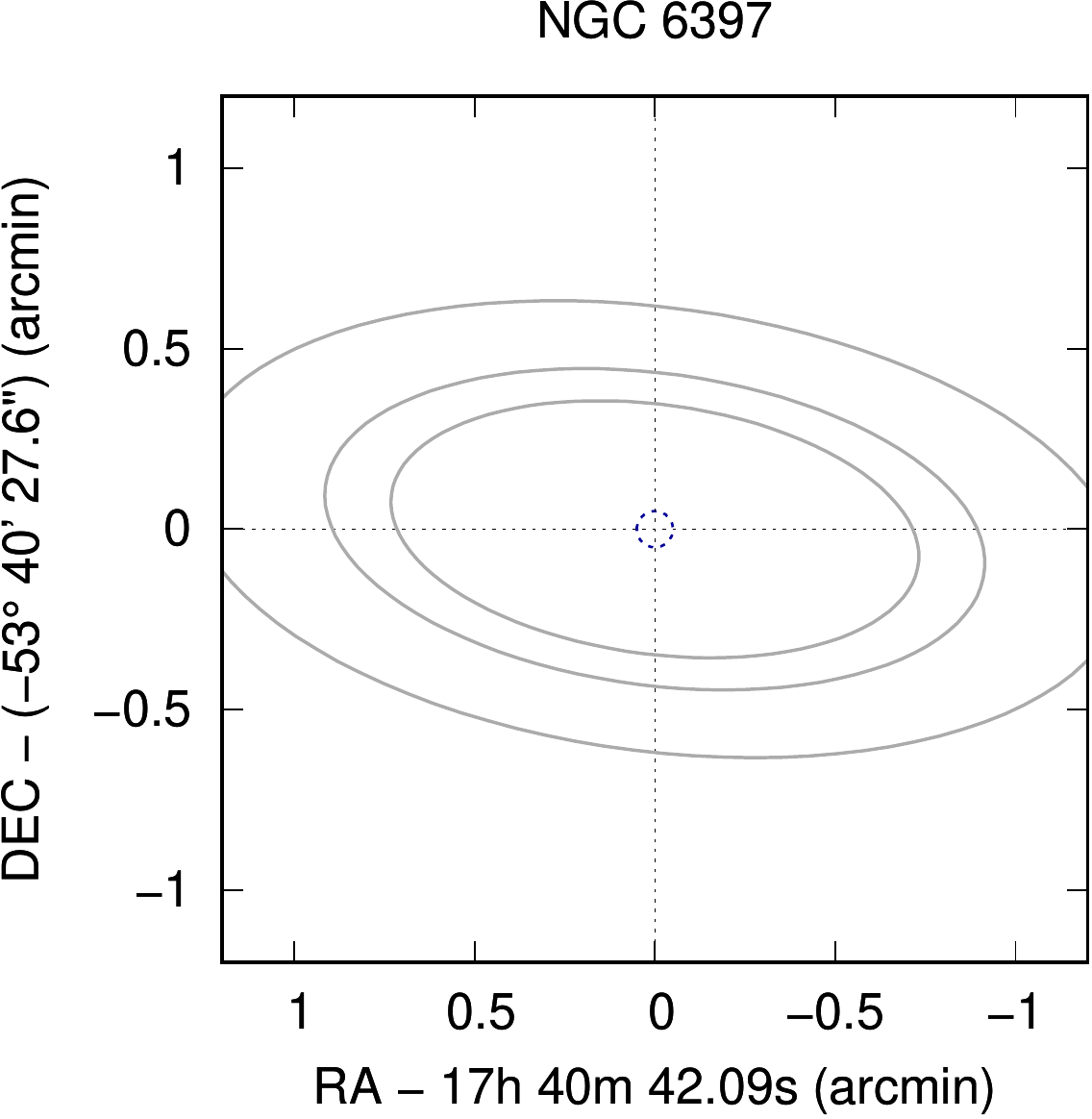}
	\\ \vskip 0.5 cm
		\includegraphics[width=0.305\textwidth]{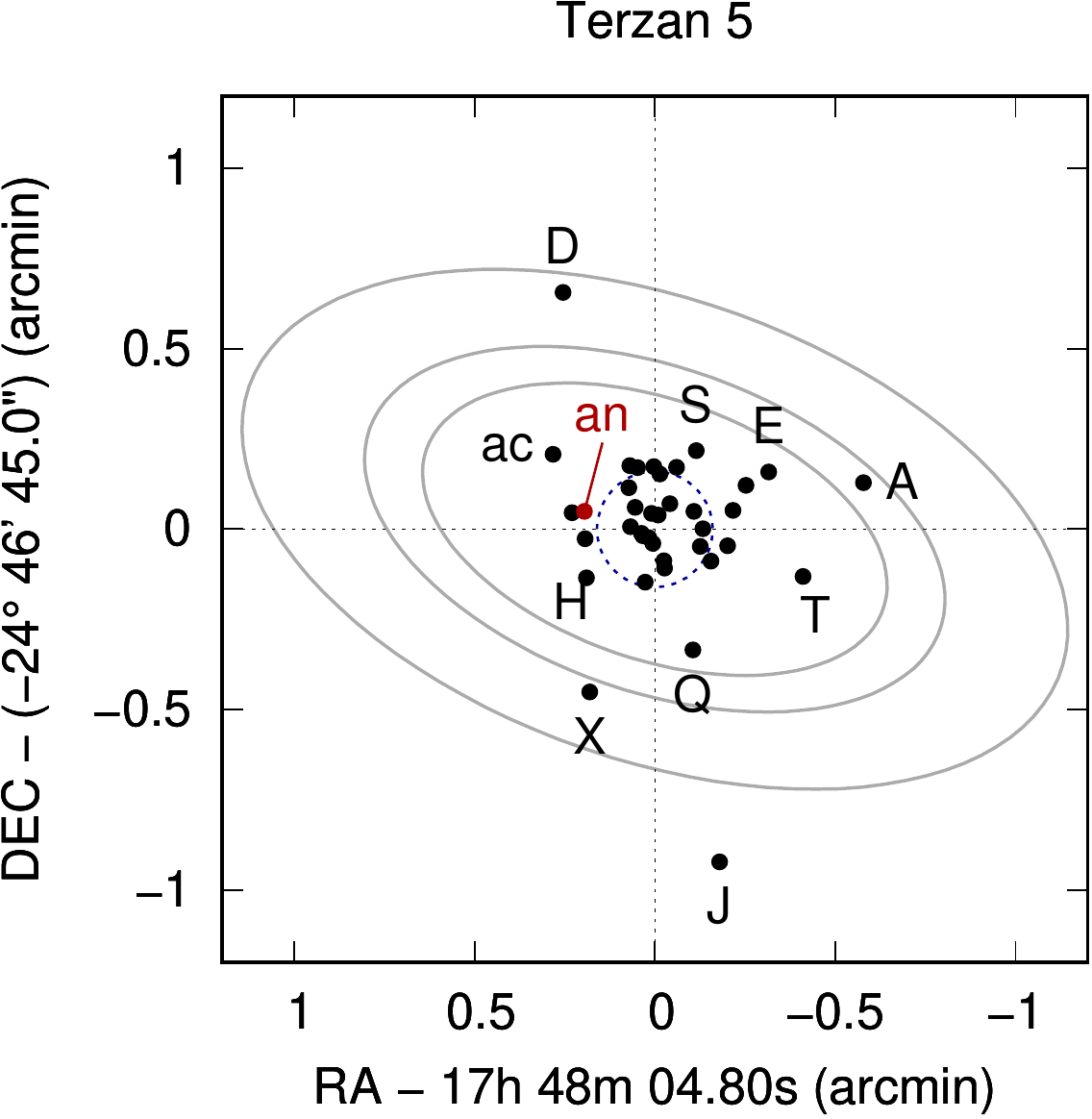}
		\qquad
		\includegraphics[width=0.305\textwidth]{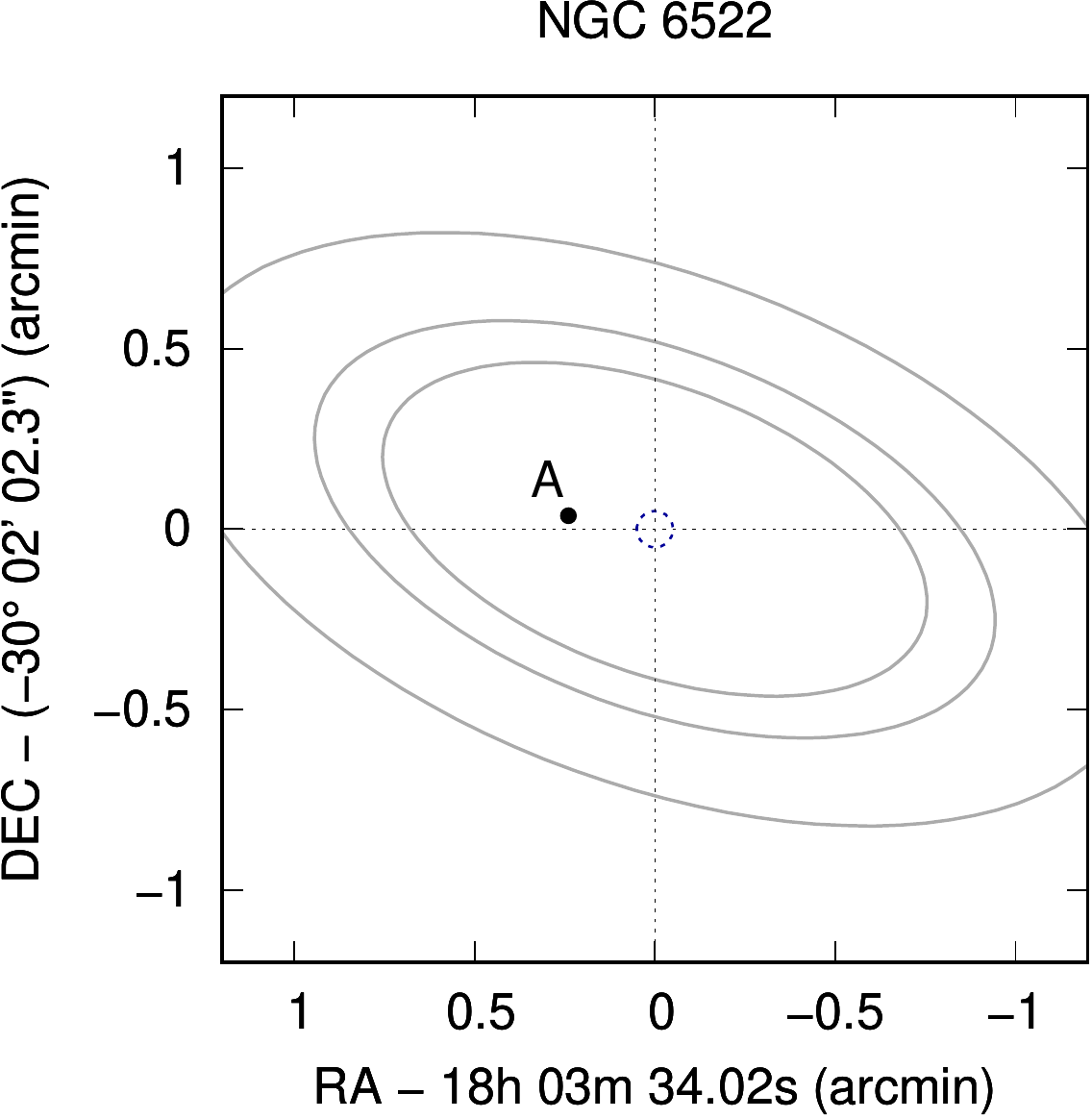}
		\qquad
		\includegraphics[width=0.305\textwidth]{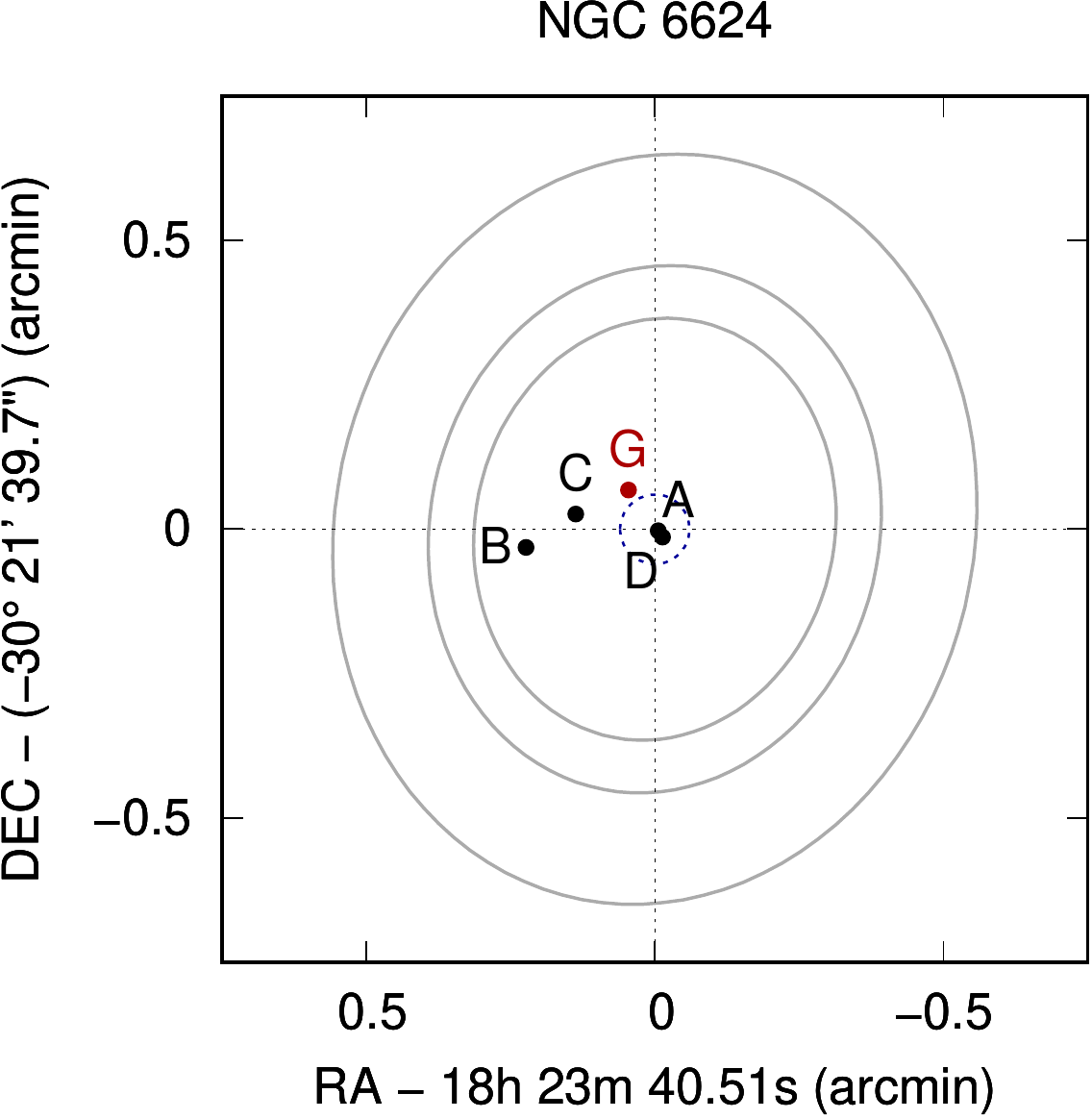}
	\\ \vskip 0.5 cm
		\includegraphics[width=0.305\textwidth]{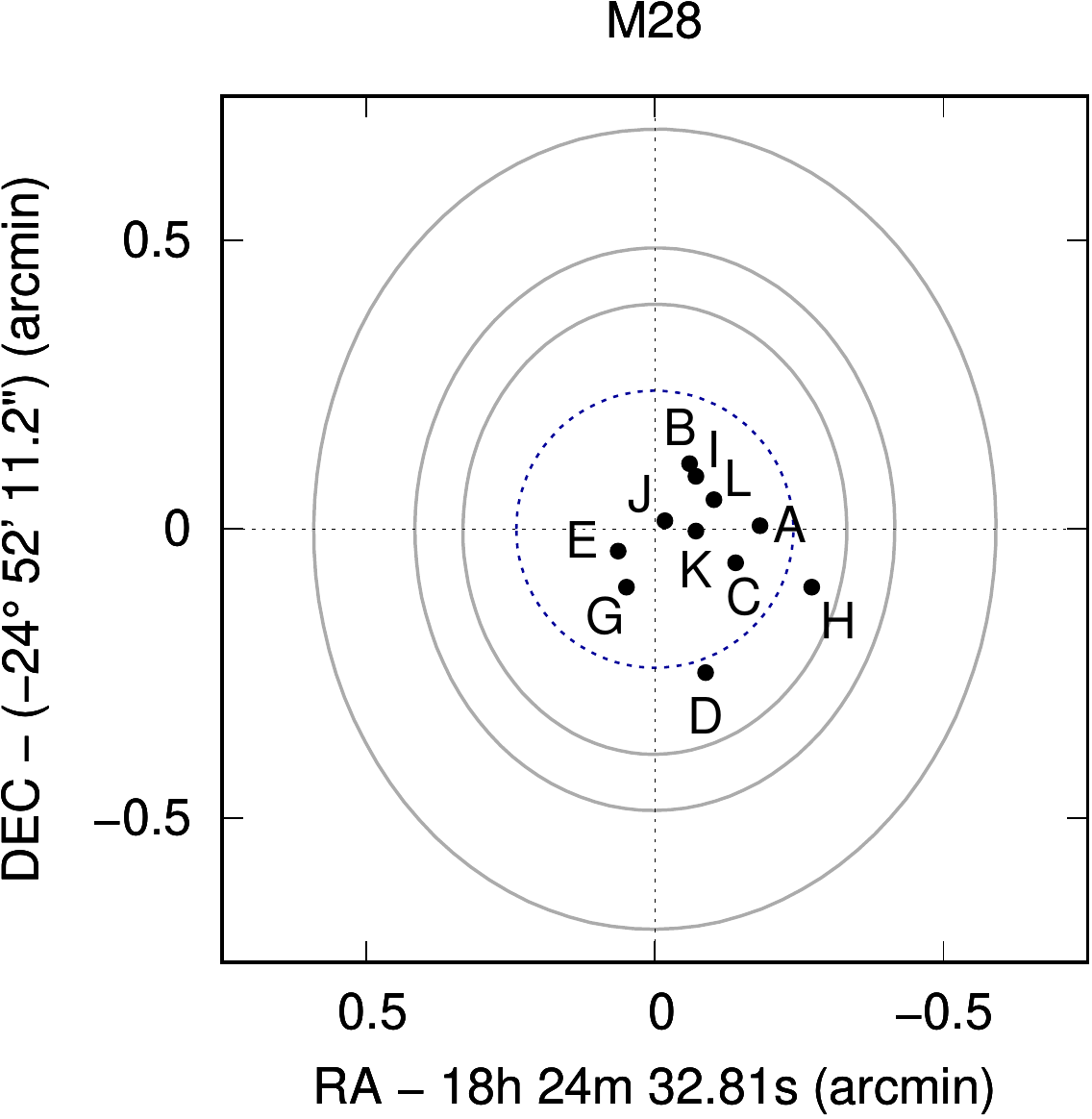}
		\qquad
		\includegraphics[width=0.305\textwidth]{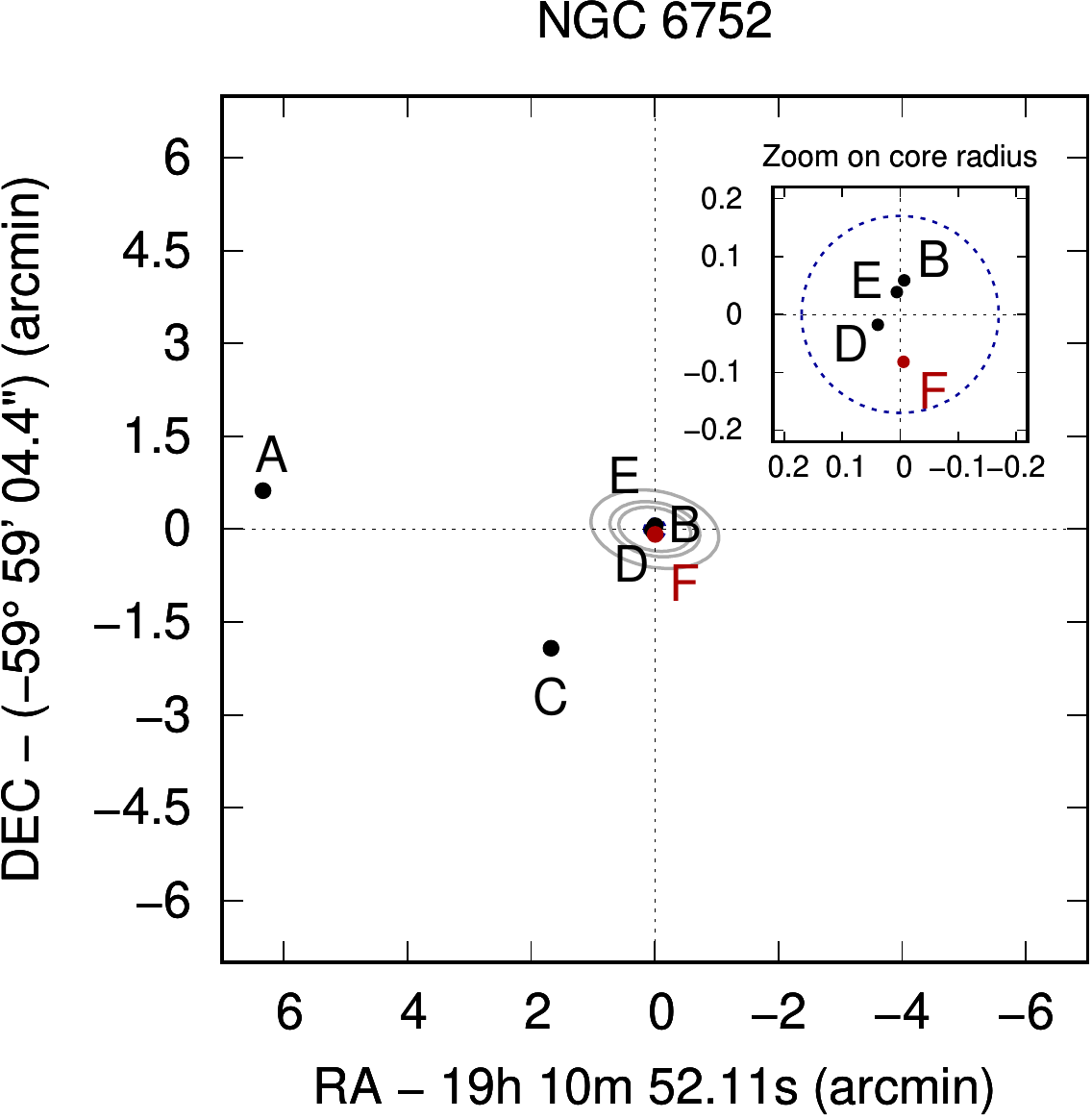}
		\qquad
		\includegraphics[width=0.305\textwidth]{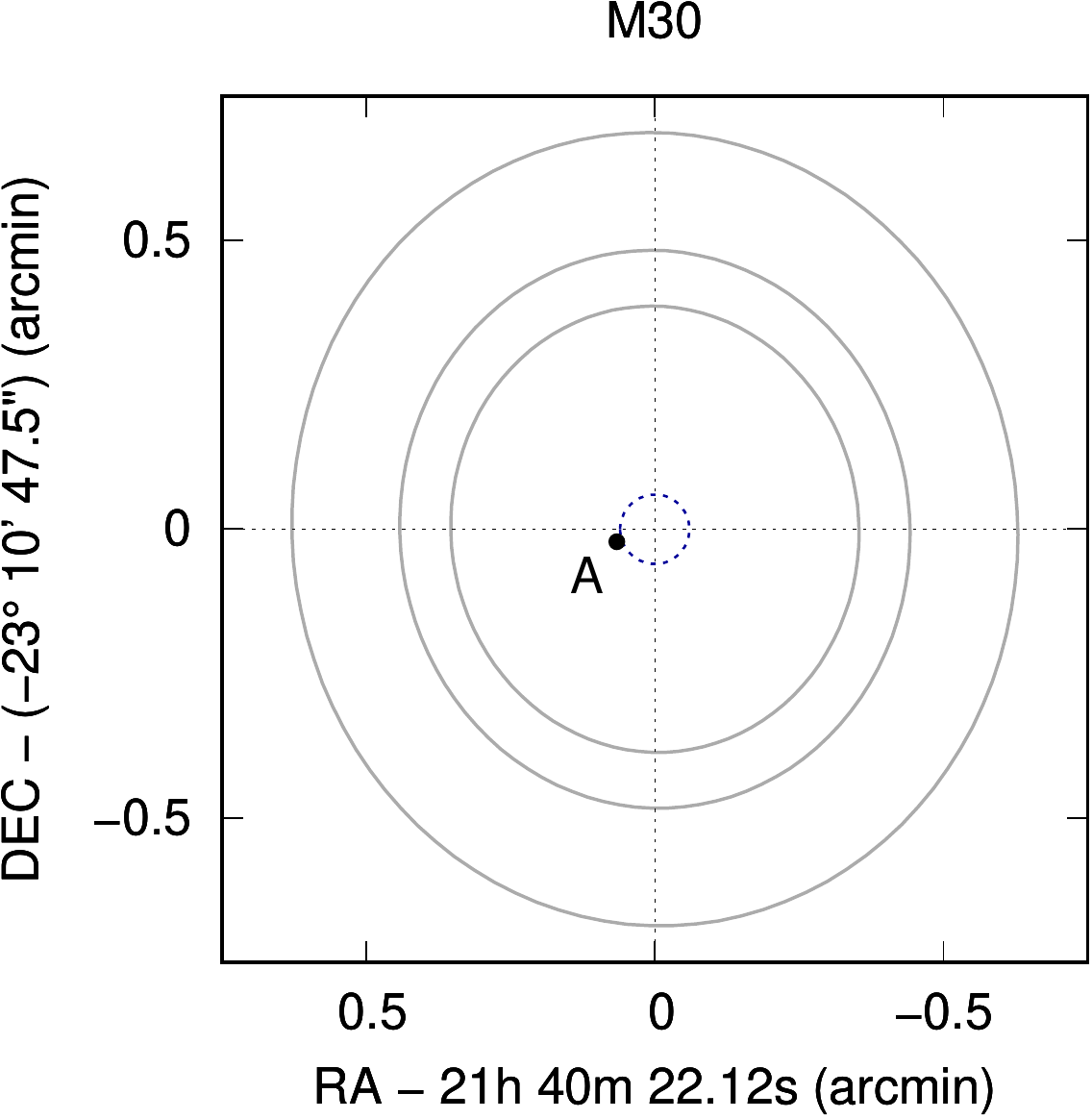}
		
  	\caption{MeerTime L-band tied array beams, shown through their contours as solid grey lines at their 50 percent level of power, for one sample observation of each of the nine GCs observed. We show the size of the beam for the top edge frequency (inner ellipse), center frequency (middle ellipse) and bottom edge frequency (outer ellipse). For each cluster, we show the core radius (dashed blue line) and all the pulsars (black dots) with a known position. }
  	\label{fig:pos_beams}
\end{figure*}

\subsubsection{47 Tuc ac, ad}
\label{sec:47Tuc_discoveries}
47 Tuc is one of the most massive GC in the Galaxy, and one of the largest in the sky. It is located at a distance of 4.69 kpc \citep{Woodley+2012}, well outside the Galactic plane, not far from the South Celestial Pole. For this reason, deep searches for radio pulsars in 47 Tuc were, until recently, only possible with the Parkes radio telescope. These searches were extremely fruitful, as they allowed the discovery of 25 pulsars in the cluster \citep{Manchester+1990,Manchester+1991,Robinson+1995,Camilo+2000,Pan+2016,Freire+2017}. Ten of these pulsars are isolated and fifteen are part of binary systems, and all are MSPs with rotation periods in the range of $2-8$ ms. 
We have discovered two previously unknown binary MSPs in this cluster, dubbed 47~Tuc~ac (PSR~J0024$-$7204ac) and 47~Tuc~ad (PSR~J0024$-$7204ad). Both of them, as all the other pulsars in 47 Tuc, are heavily affected by scintillation, which makes their follow-up challenging.

47~Tuc~ac was discovered as a 2.74-ms pulsar in the second MeerKAT observation (obs. id 02L) of the cluster, where it appeared extremely bright. It was detected at a DM of 24.46~\dmunit\ and with a line-of-sight acceleration of $a_l = -1.1$~m/s$^2$ (corresponding to a $z=11$), indicating the presence of an orbital motion affecting the observed spin period. Subsequent folding revealed the disappearance of the signal in the last fraction of the observation (rightmost panel of Figure~\ref{fig:47TucAC_folds}), likely caused by intra-binary material intercepting the pulsar's radiation, as is typical of many compact ``spiders'' \citep{Roberts2013}. No further detections of the pulsar were obtained from the other eight MeerKAT observations of 47~Tuc. To confirm the discovery, we analyzed the 16-year 47~Tuc Parkes dataset (519 observations taken in the years 1997--2013) described in \citet{Ridolfi+2016}. Doing a blind search at the DM of 47~Tuc~ac,
the pulsar was found in three different observations, in all of which the signal was occasionally eclipsed.
We used the Period-Acceleration diagram and the Periodogram methods to estimate the orbital parameters and build a first ephemeris.The latter was used to re-fold all the Parkes data, which resulted in a further two faint detections.
We then extracted ToAs from all the available detections (excluding the sections near the eclipse ingress and egress) and used \TEMPO\ to fit a simple Keplerian orbital model. However, four orbital period derivatives were needed to fit the ToAs and obtain an incoherent solution that could fold all the six detections. 47~Tuc~ac is in an orbit with $\Pb \sim 0.18$~d and  $\xp\sim0.019$~lt-s.  However, because of the meager number and quality of the detections obtained over a time span of 20 years, these parameters are not yet accurate enough to guarantee a correct orbital count.
Assuming a pulsar mass of $1.4$~\msun, the binary mass function implies a minimum companion mass $\Mc = 0.0075$~\msun, indicating that the pulsar belongs to the class of eclipsing ``black widows''.

The other newly found MSP, 47 Tuc ad, is a 3.74-ms pulsar, discovered in the fifth MeerKAT observation of the cluster (obs. id 05L) at a DM of 24.41~\dmunit. It was detected with an acceleration of  $a_l = 10.4$~m/s$^2$ ($z=30$) and some residual jerk, as well as evidence of occasional eclipses (right panel of Figure \ref{fig:47TucAD_folds}).
This is the only detection obtained from all the MeerKAT observations of the cluster. However, following the same procedure as in the case of 47~Tuc~ac, we could confirm the discovery by detecting 47~Tuc~ad in one of the Parkes observations, taken on 2004 May 27. The latter is a $7.1$-h-long pointing in which the pulsar shows up very bright in the first 1.28 h, after which the pulsed signal quickly fades away and disappears for the rest of the observation (left panel of Figure~\ref{fig:47TucAD_folds}). 
As for 47 Tuc ac, we used the Period-Acceleration diagram and Periodogram methods. We did so splitting both the MeerKAT and Parkes observations where the pulsar was detected into a few sections, so as to obtain multiple measurements of $\Pobs$ and $a_l$, in each of them. For 47~Tuc~ad, we find $\Pb = 0.3184$~d, $\xp = 0.6819$ lt-s and $\Tasc = 58831.790667$~MJD. 
The binary mass function implies a minimum companion mass of $\Mc=0.205$~\msun, hence 47~Tuc~ad is an eclipsing ``redback'' binary pulsar \citep[e.g.][]{Roberts2013}.
The range of measured values of $\Pobs$ and $a_l$ was very similar in both observations, indicating that the pulsar was detected in the same orbital phase interval. According to the binary parameters found, both detections fall in the mean anomaly orbital interval $\phib \sim 0.7-0.9$. This is not surprising, since it is on the side of the orbit closer to the observer, where pulsars in eclipsing systems are typically most likely to be detected.  Using the very long Parkes observation where 47~Tuc~ad is detected, we can also estimate the fraction of the orbit in which the pulsar is probably being eclipsed. The derived circular orbit model implies that the 7.1-h Parkes observation started at orbital phase $\phib=0.70$ and ended at $\phib=1.63$. The fact that the pulsar was detected for the first $\sim 1.28$~h means that the eclipse ingress occurred at $\phib\sim0.87$. If we assume that the eclipse is centered around the pulsar's superior conjunction ($\phib=0.25$), then the pulsar signal would be undetectable for about 76\% of the orbit, and the eclipse egress would occur at $\phib = 0.63$, i.e. exactly coinciding with when the Parkes observation ends. We remark that the assumption of the eclipse being centered at $\phib=0.25$, although reasonable, is not necessarily true, as observed for other eclipsing systems. Unfortunately, lacking a detection of 47 Tuc ad in the range $\phib = 0.63-0.70$, and scintillation being the dominant factor for the detectability of the pulsar, we cannot currently confirm this hypothesis.

47 Tuc ac and  47 Tuc ad bring the total number of pulsars known in the cluster to 27. Unfortunately, the paucity of their detections due to scintillation prevent us from obtaining phase-connected solutions, so we have no information on their spin period derivatives, nor on their exact positions (although future TRAPUM observations may help with the latter).
It is likely, however, that both pulsars are located within $\sim0.5$~arcmin from the cluster center, since this is approximately the radius, at half-power, of the tied-array beam synthesized for those observations.

\begin{table*}
\caption{Timing solutions for four newly discovered MSPs. For all solutions, the time units are TDB, the adopted terrestrial time standard is UTC(NIST) and the Solar System ephemeris used is JPL DE421 \citep{Folkner+2009}. All quoted uncertainties and upper/lower limits are 1 $\sigma$.} 
\label{tab:timing_solutions}
\begin{center}{\footnotesize
\setlength{\tabcolsep}{6pt}
\renewcommand{\arraystretch}{1.0}
\begin{tabular}{l c c c c}
\hline
\hline
Pulsar  &   M62G     & Ter 5 an            &   NGC 6624G           &  NGC 6752F                                                             \\
\hline
Alt. pulsar name    &   J1701$-$3006G                                                            &   J1748$-$2446an                                                            &   J1823$-$3021G                                                            &   J1910$-$5959F                                                            \\
\hline
Phase-connected solution?  & Partial     &    Yes          &  Yes     & Yes \\
\hline
Right Ascension, $\alpha$ (J2000)                                     \dotfill &   17:01:14.0(2)                                                          &   17:48:05.65898(16)                                                       &   18:23:40.7213(8)                                                       &   19:10:52.066(3)                                                        \\
Declination, $\delta$ (J2000)                                         \dotfill &   $-$30:06:42(41)                                                      &   $-$24:46:42.03(7)                                                      &   $-$30:21:35.63(5)                                                      &   $-$59:59:09.30(3)                                                      \\
Proper Motion in $\alpha$, $\mu_\alpha$ (mas yr$^{-1}$)               \dotfill &   --                                                                     &   $-$2.17(55)                                                              &   --                                                                     &   $-$1.9(1.4)                                                                \\
Proper Motion in $\delta$, $\mu_\delta$ (mas yr$^{-1}$)               \dotfill &   --                                                                     &   $-$28.4(18.5)                                                             &   --                                                                     &   $-$6.5(1.9)                                                                \\
Spin Frequency, $f$ (s$^{-1}$)                                        \dotfill &   217.0090551(6)                                                          &   208.23191713844(2)                                                     &   164.168745585(1)                                                       &   117.8481780116(2)                                                      \\
1st Spin Frequency derivative, $\dot{f}$ (Hz s$^{-1}$)                \dotfill &  --                                               &   $-$6.7532(3)$\times 10^{-15}$                                          &   4.8(6)$\times 10^{-16}$                                                &   $-$1.02942(9)$\times 10^{-14}$                                         \\
2nd Spin Frequency derivative, $\ddot{f}$ (Hz s$^{-2}$)               \dotfill &   --                                                                     &   $-$8.83(2)$\times 10^{-25}$                                            &   --                                                                     &   $-$9.37(24)$\times 10^{-26}$                                             \\

3rd Spin Frequency derivative, $\ddot{f}$ (Hz s$^{-3}$)               \dotfill &   --                                                                     &   $-$4.586(56)$\times 10^{-33}$                                            &   --                                                                     &   --                                             \\

Reference Epoch (MJD)                                                 \dotfill &   58700.000                                                           &   56000.000                                                           &   58700.000                                                           &   58700.000                                                           \\
Start of Timing Data (MJD)                                            \dotfill &   58602.827                                                              &   53204.042                                                              &   58736.722                                                              &   51468.265                                                              \\
End of Timing Data (MJD)                                              \dotfill &   59074.864                                                              &   58933.449                                                              &   59072.968                                                              &   58850.722                                                              \\
Dispersion Measure, DM (pc cm$^{-3}$)                                 \dotfill &   113.679(7)                                                             &   237.736(8)                                                             &   86.206(6)                                                              &   33.20(3)                                                               \\
Number of ToAs                                                        \dotfill &   69                                                                  &   198                                                                &   125                                                                &   67                                                                  \\
Residuals RMS ($\mu$s)                                                \dotfill &   19.15                                                                  &   29.17                                                                  &   18.55                                                                  &   38.25                                                                  \\
\hline
\multicolumn{5}{c}{Binary Parameters}  \\
\hline\hline
Binary Model                                                          \dotfill &   ELL1                                                                   &   DD                                                                     &   DD                                                                     &   --                                                                     \\
Projected Semi-major Axis, $x_{\rm p}$ (lt-s)                         \dotfill &   0.620316(8)                                                            &   12.7820385(35)                                                           &   3.00331(2)                                                             &   --                                                                     \\
Orbital Period, $P_b$ (d)                                          \dotfill &   0.77443355(5)                                                          &   9.6197533(8)                                                           &   1.54013654(5)                                                          &   --                                                                     \\
Orbital Eccentricity, $e$                                             \dotfill &   --                                                                     &   0.0065856(6)                                                           &   0.380466(6)                                                            &   --                                                                     \\
Epoch of Periastron, $T_0$ (MJD)                           \dotfill &   --                                                                     &   56743.6892(1)                                                          &   58909.681498(8)                                                        &   --                                                                     \\
Longitude of Periastron, $\omega$ (deg)                               \dotfill &   --                                                                     &   $-$145.2624(45)                                                          &   146.721(1)                                                             &   --                                                                     \\
First Laplace-Lagrange parameter, $\epsilon_1$                              \dotfill &   6.08(28)$\times 10^{-4}$                                                 &   --                                                                     &   --                                                                     &   --                                                                     \\
First Laplace-Lagrange parameter, $\epsilon_2$                            \dotfill &   6.73(24)$\times 10^{-4}$                                                 &   --                                                                     &   --                                                                     &   --                                                                     \\
Epoch of Ascending Node, $T_\textrm{asc}$ (MJD)            \dotfill &   58894.966321(3)                                                        &   --                                                                     &   --                                                                     &   --                                                                     \\
Rate of periastron advance, $\dot{\omega}$ (deg\,yr$^{-1}$)                   \dotfill &   --                                                                     &   0.0095(11)                                                               &   0.217(4)                                                               &   --                                                                     \\
Orbital Period derivative, $\dot{P}_{\rm b}$ (s s$^{-1}$)  \dotfill &   --                                                                     &   26.8(3.4)~$\times 10^{-12}$                                                                   &   --                                                                     &   --                                                                     \\
\hline
\multicolumn{5}{c}{Derived Parameters}  \\
\hline\hline
Spin Period, $P$ (ms)                                                  \dotfill &   4.60810264(1)                                                       &   4.8023377671500(4)                                                  &   6.09129342145(4)                                                    &   8.48549393697(1)                                                    \\
1st Spin Period derivative, $\dot{P}$ (s s$^{-1}$)                    \dotfill &   --                                                &   1.55746(6)$\times 10^{-19}$                                            &   $-$1.8(2)$\times 10^{-20}$                                             &   7.4122(6)$\times 10^{-19}$                                             \\
Mass Function, $f(M_{\rm p})$ (\msun)                         \dotfill &   4.27$\times 10^{-4}$                                                   &   0.02                                                                   &   0.01                                                                   &   --                                                                     \\
Minimum companion mass, $M_{\rm c, min}$ (\msun)              \dotfill &   0.10                                                                   &   0.43                                                                   &   0.33                                                                   &   --                                                                     \\
Median companion mass, $M_{\rm c, med}$ (\msun)             \dotfill &   0.11                                                                   &   0.52                                                                   &   0.39                                                                   &   --                                                                     \\
Total System Mass (assuming GR), $M_{\rm tot}$ (\msun)                                           \dotfill &   --                                                                     &   2.97(52)                                                                  &   2.65(7)                                                                   &   --                                                                     \\
Intrinsic Spin-down, $\dot{P}_{\rm int}$ (s s$^{-1}$)      \dotfill &   --                                                                     &   8.49(2)$\times 10^{-22}$                                                                  &   --                                                                     &   --                                                                     \\
Surface Magnetic Field, $B_0$, (G)                           \dotfill &   --                                                                     &   $< 3.12 \times 10^{8\phantom{0}}$                                                                   &   --                                                                     &   --                                                                     \\
Spin-down Luminosity, $L_{\rm sd}$ (erg s$^{-1}$) \dotfill &   --                                                                     &   $< 7.24 \times 10^{33}$                                                                   &   --                                                                     &   --                                                                     \\
Characteristic Age, $\tau_{\rm c}$ (Gyr)                              \dotfill &   --                                                                     &   $> 3.75$                                                                  &   --                                                                     &   --                                                                     \\
Offset from GC center in $\alpha$, $\theta_\alpha$                    \dotfill &   --                                                                  &   0.195                                                                  &   0.046                                                                  &   $-$0.006                                                               \\
Offset from GC center in $\delta$, $\theta_\delta$                    \dotfill &   --                                                                  &   0.050                                                                  &   0.068                                                                  &   $-$0.082                                                               \\
Total offset from GC center, $\theta_\perp$ (arcmin)                  \dotfill &   --                                                                  &   0.201                                                                  &   0.082                                                                  &   0.082                                                                  \\
Proj. distance from GC center, $r_\perp$ (pc)                         \dotfill &   --                                                                  &   0.404                                                                  &   0.188                                                                  &   0.095                                                                  \\
Proj. distance from GC center, $r_\perp$ (core radii)                 \dotfill &   --                                                                  &   1.257                                                                  &   1.363                                                                  &   0.482                                                                  \\
\hline
\end{tabular} }
\end{center} 
\end{table*}

\subsubsection{M62G}
M62 (also known as NGC 6266), is a massive cluster located at a distance of 6.8 kpc in the direction of the Galactic bulge. Although previous estimates classified this cluster as core-collapsed \citep{Djorgovski_Meylan1993}, more recent studies rule out this hypothesis \citep{Beccari+2006}. This is supported by the observed pulsar population of M62: all the six known MSPs are circular binaries \citep{Damico+2001a,Possenti+2003,Lynch+2012} with characteristics that are typical of unperturbed binary evolution.

We have found a new 4.61-ms binary MSP,  named M62G (or PSR~J1701$-$3006G). It was discovered at a ${\rm DM}=113.68$~\dmunit\ in the lower half of the band of the first, $\sim$1~h MeerKAT observation (obs. id 01L), with an acceleration of $\simeq 1.5$~m/s$^{-2}$, hinting at a steep-spectrum binary pulsar (although this could also be an observational bias, if the pulsar is far from the tied-array beam boresight). M62G was re-detected in all the subsequent observations of the cluster (which included a dense orbital campaign consisting of five observations taken within 10 days), always with nominal absolute line-of-sight accelerations ranging from $0.5-1.6$~m/s$^{-2}$, confirming the binary nature. Using the Period-Acceleration method first, and the $P_{\rm obs}(t)$ fitting of the orbital campaign observations, we derive a circular binary orbit with $\Pb \sim  0.77$~d and $\xp = 0.62$~lt-s, implying a companion of minimum mass  $\Mc = 0.099$~\msun\ (as always, assuming $\Mp = 1.4$~\msun). Pulsar G was observed and detected by MeerKAT at all possible orbital phases, without any signs of eclipses. Hence it is likely that the companion star is a white dwarf (WD). 
Using all the MeerKAT observations, we computed the ToAs, and fitted the latter with \TEMPO. Unfortunately, the detections are too sparse to guarantee a phase-coherent timing solution throughout the dataset. However, we could achieve unambiguous phase connection on part of the data. This was enough to reveal that the eccentricity of the system is very low ($e=9.1\times 10^{-4}$) but non-zero, leading to a high covariance between the longitude of periastron, $\omega$, and the epoch of periastron passage, $T_0$. For this reason, we opted for an ELL1 binary model, which fits for $\xp$, $\Pb$ and $\Tasc$ and for the so-called Laplace-Lagrange parameters ($\epsilon_1 = e  \, \sin\omega$ and $\epsilon_2 = e \, \cos\omega$, \citealt{Lange+2001}).
The parameters derived with the partially connected solution for this pulsar are reported in Table~\ref{tab:timing_solutions}.

\subsubsection{Ter 5 an}
Ter 5 is a GC located well within the Galactic bulge. With 38 radio pulsars known \citep{Lyne+1990,Ransom2001,Ransom+2005,Cadelano+2018,Andersen_Ransom2018}, it holds the record for the highest number of pulsars hosted, and accounts for about a quarter of the total GC pulsar population. 

\begin{figure*}
\begin{center}
    \includegraphics[width=0.87\textwidth]{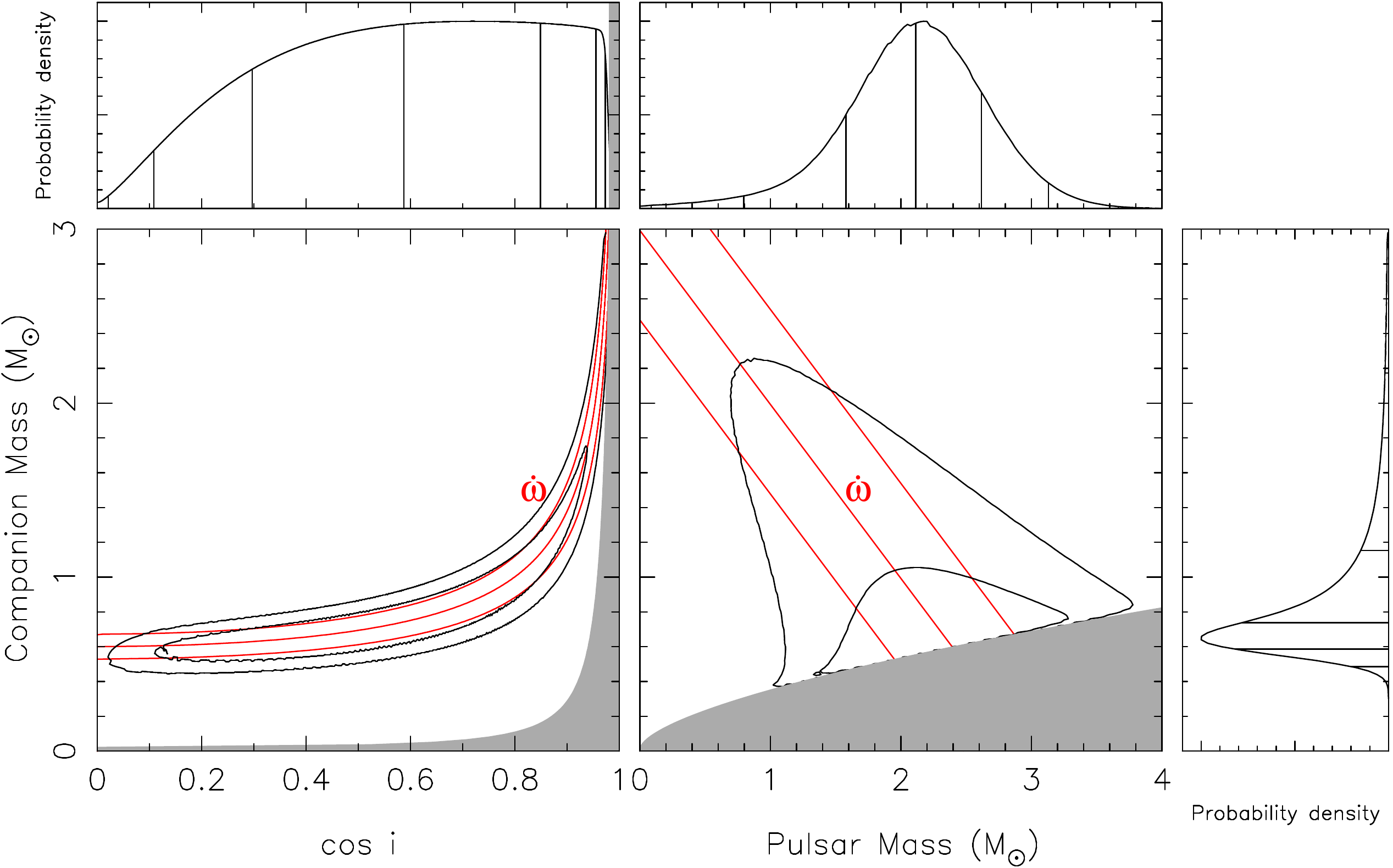}\vskip 0.5 cm
    \includegraphics[width=0.87\textwidth]{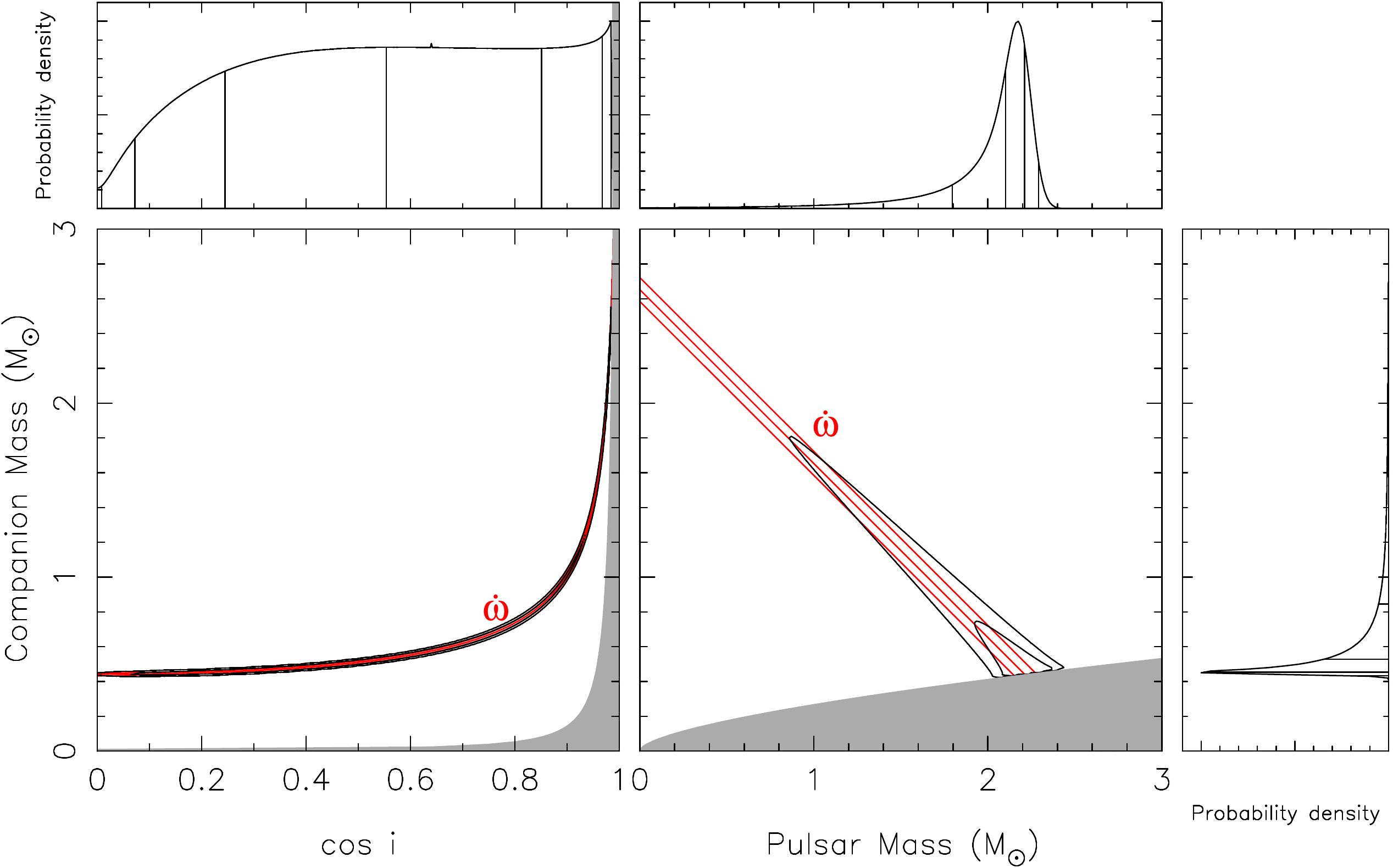}
\caption{Mass-inclination and mass-mass diagrams for Ter 5 an (top) and NGC 6624G (bottom). The main square panels depict 
the $\cos i$-$\Mc$ and $\Mc$-$\Mp$ planes. The grey areas in the former are excluded by the requirement that the mass of the pulsar is positive, the grey areas in the latter are excluded by the mass function and the requirement that $\sin i \leq  1$. The red lines depict the masses consistent with the measurement of $\dot{\omega}$ and its $\pm  1$-$\sigma$  uncertainties, under the assumption that this effect is dominated by the GR contribution and that GR is the correct theory of gravity.  The contours include 68.3 and 95.4\% of a 2-D probability distribution function (pdf) derived from the $\chi^2$ of \TEMPO\ fits that assumed all GR effects to be according to the masses and orbital inclination at each point. The side panels show the probability density functions for the $\cos i$ (top left), $\Mp$ (top right) and $\Mc$ (right) derived by marginalizing the aforementioned 2-D pdfs. For Ter~5~an, the estimated median pulsar mass is $2.13$~\msun\ and median companion mass is $0.75$~\msun. In the case of NGC 6624G, we obtain a median for the pulsar mass of $2.1$~\msun, but with a tail of probability  that extends to lower masses: there is a 31\% probability of $\Mp < 2$~\msun\ and a 3.8\% probability of $\Mp < 1.4$~\msun. Thus, either the pulsar is very massive, or it has a massive companion; the system was likely formed in an exchange encounter. 
The timing for both systems shows no evidence for the Shapiro delay nor any other relativistic effect that might allow a determination  of the individual masses, as we can see from the fact that only high inclinations
(near $\cos i \, = \, 0$) are excluded by the lack of a detectable Shapiro delay. Apart from these, the pdf is very flat in $\cos i$.
}
\label{fig:masses}
\end{center}
\end{figure*}

In this cluster, we have discovered a new 4.80-ms binary MSP, at a DM of 237.74~\dmunit. The pulsar, called Ter~5~an (PSR~J1748$-$2446an), was detected in the first MeerKAT observation of the cluster (obs. id 01L) with a small acceleration of $+0.222$~m/s$^2$. The period and acceleration matched those of a candidate found in the initial set of searches performed in 2005 on the GBT data, thereby instantly confirming it. To determine the orbital parameters, a dense orbital campaign of four 3.5-h-long observations was carried out on the pulsar over the course of three days. These observations covered about a quarter of the orbit, enough to fit $\Pobs(t)$ and have a first estimate of $\Pb$ and $\xp$. Using this initial orbital model, we could fold and detect the pulsar in a number of archival GBT observations of Ter~5, taken with the Spigot \citep{Kaplan+2005} and GUPPI \citep{DuPlain+2007} backends with the L-band and S-band receivers. These were used to further refine the orbital parameters: the pulsar was found to be in a $\sim9.62$-day, slightly eccentric orbit. With more certain binary parameters, the pulsar could then be detected in a large number (63) of observations in the archival data, providing us with a total timing baseline of about 16 years (2004--2020).
Combining the MeerKAT and the GBT detections, we obtained a phase-connected timing solution using the theory-independent DD binary model. Ter~5~an is in a binary orbit with very likely a WD companion of minimum mass of 0.43~\msun\ (for $\Mp = 1.4$~\msun). The orbital eccentricity is $e=0.0066$, and it is likely due to the stellar interactions encountered by the system in the dense environment of the cluster. We were also able to measure the rate of advance of the periastron, which is  $\omegadot = 0.009 \pm 0.001$~deg\,yr$^{-1}$. Assuming that this is fully relativistic, and that General Relativity (GR) is the correct theory of gravity, this translates into a total mass of the system $\Mtot= 2.97 \pm 0.52$~\msun. 
Taking the uncertainty of $\dot{\omega}$ into account, and assuming a priori randomly aligned orbits, we made a $\chi^2$ map to constrain the component masses of the system.
We find that the pulsar mass is $\Mp = 2.13^{+1.01}_{-1.32}$~\msun\ and the companion mass is $\Mc~=~0.75^{+1.38}_{-0.25}$~\msun\ at the $2 \sigma$ confidence level. Although the constraints on the masses are very loose, we can at least exclude very high orbital inclinations, given that $i < 83.76^\circ$ at the $2\sigma$ level.
The resulting mass-mass diagram obtained for Ter~5~an is shown in Figure~\ref{fig:masses}. We have also measured the orbital period derivative, $\dot{\Pb} = 26.8 \pm 3.4 \times 10^{-12}$ which, in the case of GC pulsars, includes contributions from accelerations due to the cluster potential, Galactic potential and Shklovskii effect (caused by the proper motion of the system, \citealt{Shklovskii1970}) combined.
The long timing baseline has also allowed us to measure the pulsar astrometric and kinematic parameters. Ter~5~an is located $\sim 0.2$ arcmin (1.26 core radii) east of the cluster center (Figure~\ref{fig:pos_beams}). The proper motion is still loosely constrained, with measured values of $\mu_\alpha=-2.1\pm0.6$~\pmunit\ and $\mu_\delta=-28 \pm 18$~\pmunit, consistent with those of the other pulsars in the cluster (S. Ransom, private comm.).
As for the rotational parameters, we measured up to three spin frequency derivatives for Ter 5 an. The first derivative is due to the intrinsic spin-down, as well as the cluster gravitational potential. The second derivative likely has contributions from both the cluster potential and the presence of nearby stars. Higher derivatives are entirely due to local potentials around the system in the cluster, as has been studied with other known pulsars in Ter~5 \citep{Prager+2017}.
Since the acceleration contributions to both spin and orbital period derivatives are similar, the intrinsic spin period derivative, $\dot P_{\rm int}$, can be calculated from the relation $\Pdotint = \Pdotobs - P \, \Pbdotobs / \Pb$ (considering $\dot P_{{\rm b, int}}\sim0$, which is the case for wide orbits with negligible gravitational wave damping). We obtain $ \dot P_{\rm int}~=~(8.49~\pm~0.02)~\times~10^{-22}$. From this, we can constrain the surface magnetic field, $B_{\rm s} \leq 3.1 \times 10^8$~G, and characteristic age of the pulsar, $\tau_{\rm c} \geq 3.75$~Gyr, at the 1-$\sigma$ level. 

\subsubsection{NGC 6522D}
NGC 6522 is a  cluster located at a distance of 7.7 kpc, close to the Galactic center. It is core-collapsed and hosts three previously known MSPs, all of which are isolated \citep{Possenti+2005,Begin2006,Zhang+2020}. We have discovered a fourth isolated MSP, named NGC 6522D (or PSR~J1803$-$3002D), with a spin period of 5.53 ms and a DM of 192.73~\dmunit. The pulsar was found in the first L-band MeerKAT observation of the cluster (obs. id 01L) with no acceleration. It was then re-detected, via direct folding with the nominal barycentric period, in the second L-band observation and in a third, 40-min-long follow-up observation made in the UHF band.
We have also searched for NGC 6522D in 154 L-band observations taken in the years 2000--2012 with the Parkes telescope in the context of the Parkes Globular Cluster Survey \citep[PKSGC,][]{Possenti+2005}. However, we were unable to detect the pulsar in these data. This is not surprising, given the significantly lower sensitivity of the Parkes telescope and the high DM of the cluster, which does not allow scintillation to occasionally boost the pulsar signal over the detectability threshold, as happens for other clusters. Follow-up timing with high-gain telescopes like MeerKAT or the GBT is therefore required to obtain the astrometric and kinematic parameters of NGC 6522D.

\begin{figure}
\centering
	\includegraphics[width=\columnwidth]{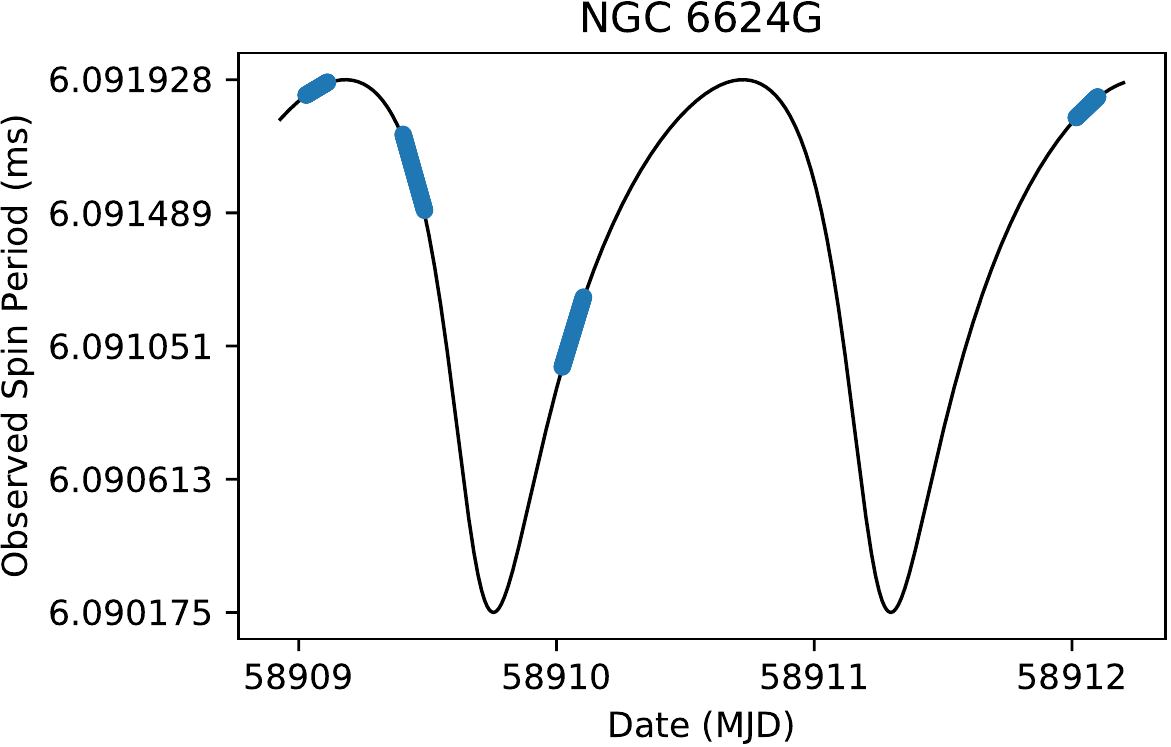}
  	\caption{{\em Blue dots}: observed spin period of NGC 6624G as a function of time, as measured during the four orbital campaign observations of the cluster; these clearly show that the pulsar orbit has a significant eccentricity.
  	  {\em Black line}: best-fitting orbital model ($\Pb = 1.54$~d, $\xp=3.00$~lt-s, $e=0.38$,  $\omega=146.7$~deg)}
  	\label{fig:NGC6624G_orbit}
\end{figure}

\subsubsection{NGC 6624G, H}
\label{sec:discoveries_NGC6624}
NGC 6624 is another core-collapsed cluster, known to be the likely host of an intermediate-mass black hole at its center \citep{Perera+2017}. Its radio pulsar population is composed of one binary and five isolated pulsars. Two of the isolated pulsars are unrecycled ones with spin periods > 300 ms, whereas the others are all MSPs \citep{Lynch+2012} .

In this cluster, we have found two new MSPs, NGC~6624G (PSR~J1823$-$3021G) and NGC~6624H (PSR~J1823$-$3021H).

NGC 6624G is a 6.09-ms pulsar in a highly eccentric binary orbit. The pulsar was discovered in the full-length segment (2.5 h) of the first observation (obs. id 01L) of the cluster, with a line-of-sight acceleration of $-0.238$ m/s$^2$ ($z=11$) and some leftover jerk in the signal. The pulsar was also re-detected in the four successive survey observations, which enabled us to use the Period-Acceleration diagram method to estimate the orbital parameters. The diagram indicated a relatively wide orbit, with signs of a non-zero eccentricity. To better characterize the orbit, we conducted a dense orbital campaign consisting of four, 2-h-long observations over a time span of about three days, in each of which NGC 6624G was easily detected. A fit of $\Pobs(t)$ using these observations revealed a 1.54-day orbit with an eccentricity of $e=0.38$ (Figure \ref{fig:NGC6624G_orbit}). The pulsar orbit has a projected semi-major axis of $\xp=3.00$~lt-s, and a longitude of periastron $\omega=146.7$~deg. 
The 13 MeerKAT observations of the cluster enabled the determination of a phase-connected timing solution for NGC~6624G, over a time span of $\sim11$~months, which is listed in Table \ref{tab:timing_solutions}. For the latter, we used the DD binary model, as in the case of Ter~5~an. 
The time spanned by the solution allowed us to have first estimates of the pulsar position and its $\Pdot$. The timing position places NGC~6624G at $\sim0.08$~arcmin, roughly in the North-East direction, from the nominal cluster center. As we shall see in Section \ref{sec:localization}, this position is confirmed by the TRAPUM observations of the cluster.
The observed spin period derivative is $\Pdot=(-1.8\pm 0.2)\times10^{-20}$. From the negative value, we deduce that the line-of-sight acceleration caused by the cluster's gravitational field is significant in NGC 6624G, and the latter must be located on the far side of the cluster. This seems to be at odds with the measured DM of $86.206\pm0.06$~\dmunit, which is the lowest among all the known pulsars in NGC~6624. This means that the DM variations are likely dominated by the Galactic foreground, as in nearly all known GCs, and not by gas within the cluster, as in the case of 47~Tuc \citep{Freire+2001b}.
Thanks to the high eccentricity of the system, 11 months of timing were also sufficient to measure the rate of advance of periastron to high significance: $\omegadot = 0.217\pm 0.004$~deg\,yr$^{-1}$. Assuming that this is purely relativistic and that
GR is the correct theory of gravity, this implies that the total mass of the system is $\Mtot = 2.65\pm0.07$~\msun. Also, the $\omegadot$ measurement constrains the minimum mass of the companion to $\Mc \gtrsim 0.44$~\msun, the maximum pulsar mass 
$\Mp \lesssim 2.4$~\msun,
and the system inclination $i\gtrsim 9.6$~deg (see Figure \ref{fig:masses}). 
As in the case of Ter 5 an, we made a $\chi^2$ map to constrain the component masses of the system, considering the uncertainty of $\dot{\omega}$, and assuming a priori randomly aligned orbits. For this system, we find that the pulsar mass is $\Mp = 2.10^{+0.19}_{-1.23}$~\msun\ and the companion mass is $\Mc = 0.53^{+1.30}_{-0.09}$~\msun\ at the $2 \sigma$ confidence level.
Even though the masses are still loosely constrained, the probability that the pulsar has a mass $>$ 2~\msun\ is 69 per cent. The detection of a second relativistic effect will be necessary to have precise measurements of the individual masses, but no such effects have been detected in the timing so far.
Finally, we note that the companion star could be either a massive WD or a NS. Regardless of its nature, the current companion is too massive to be the remnant of the star that recycled the pulsar: NGC 6624G is very likely the result of a ``secondary'' exchange encounter, where the original companion that spun the pulsar up was ejected and replaced by the current companion. Comparing it to the other secondary exchange products mentioned above, we find that contrary to M15C, and similarly to NGC~1851A, NGC 6624G is very close to the cluster center. This is very likely to be the result of mass segregation: being much more massive than the average stellar population in NGC 6624, dynamical friction caused the system to quickly sink towards the center.

The second new pulsar in the cluster, NGC 6624H, is isolated and has a spin period of 5.13 ms. It was discovered with no signs of acceleration in the fourth observation of the cluster (obs. id 04L), at a DM of 86.85~\dmunit, and subsequently re-detected, through folding, in most of the other observations of NGC 6624. Because of its faintness, we were able to phase connect only a few groups of ToAs, but were unable to derive a coherent solution across all the data. Nevertheless, thanks to the TRAPUM observations of the cluster, we know that pulsar H is most likely located at less than $\sim 0.25$ arcmin from the center of NGC 6624 (see Section \ref{sec:localization}).

\subsubsection{NGC~6752F}
NGC 6752 is a core-collapsed cluster hosting five known pulsars.
The peculiarity of this cluster is the location of two of these pulsars, namely pulsar A (a relativistic pulsar-WD binary, \citealt{Damico+2001a, Corongiu+2006,Corongiu+2012}) and C (an isolated MSP, \citealt{Damico+2001b}), both of which are extremely far from the cluster center. The other three pulsars B, D and E, which are also isolated MSPs, \citep{Damico+2001b, Damico+2002} are much closer to the cluster center, a natural consequence of mass segregation \citep{Ferraro+1997}.

We have discovered a new 8.48-ms isolated MSP, dubbed NGC~6752F (or PSR J1910$-$5959F). The pulsar was found in the second, 2.5-h-long observation (obs id. 02L) of the cluster, at a DM of 33.22~\dmunit\ and with no measurable acceleration. We then revisited the preceding observation of NGC~6752 (obs id. 01L) and obtained a second, fainter detection, again with no evidence of acceleration, confirming the isolated nature of the pulsar. Searching for the same periodicity in archival L-band data of the cluster taken at Parkes in the interval 1999--2016, we were able to re-detect pulsar F in 40 other observations. Although in a handful of these NGC~6752F showed up fairly brightly, the vast majority of the detections were extremely faint. The elusiveness of the pulsar in the Parkes dataset is due to its intrinsic faintness and to the strong scintillation that affects the pulsars of this cluster, which only occasionally brings the flux density of NGC 6752F above the detection threshold of Parkes.
Combining the ToAs obtained from the Parkes and MeerKAT detections, we were able to obtain a phase-connected timing solution spanning $\sim20$ years (1999--2020), which is listed in Table \ref{tab:timing_solutions}. The pulsar is located within the core radius of the cluster, about  0.1 arcmin South of the nominal cluster center (Figure \ref{fig:pos_beams}).  The long time spanned by the data also enabled the detection of two spin frequency derivatives, likely due to the cluster's and local stellar gravitational potentials experienced by the pulsar.
As the first spin period derivative is positive ($\Pdot = 7.4\times10^{-19}$), we cannot say whether NGC 6752F is located on the near or far side of the cluster, nor can we use the pulsar to put additional constrain to the mass-to-light ratio of this cluster \citep{Damico+2002}.
We were also able to measure the proper motion of the pulsar in both Right Ascension and Declination, which we find to be $\mu_\alpha=-1.9\pm1.4$~\pmunit and $\mu_\delta=-6.5 \pm 1.9$~\pmunit, respectively. Considering a distance to the cluster of 4.0 kpc \citep{Harris2010}, the total proper motion of $\mu_{\rm tot}=6.80\pm2.17$~\pmunit\ implies that the pulsar is moving with a transverse velocity of $\sim 129$~km\,s$^{-1}$ .

The bulk proper motion of NGC~6752 has recently been measured by \citet{GaiaCollaboration+2018}: $\mu_\alpha = -3.1908 \pm 0.0018$~\pmunit and $\mu_\delta = -4.0347 \pm 0.0020$~\pmunit. This imply that the pulsar proper motion relative to the cluster is $\Delta\mu_\alpha \sim 1.3 \pm 1.4 $~\pmunit\ and $\Delta\mu_\delta \sim -2.5 \pm 1.9 $~\pmunit\, corresponding to a relative transverse velocity in the range $19-87$~km/s, at the 1-$\sigma$ level. Although the uncertainty is large, this is still consistent with the pulsar being bound to the cluster, as the latter has a central escape velocity of $\sim30$~km/s \citep{Colpi+2003}.

\subsection{Characterization of previously known pulsars}
\label{sec:previously_known_pulsars}
Apart from the search and discovery of new pulsars, the first MeerKAT GC census observations are valuable for a number of additional scientific studies. Here we report a few more results on previously known pulsars, obtained using the collected data.

\begin{figure}
\centering
    \includegraphics[width=0.44\columnwidth]{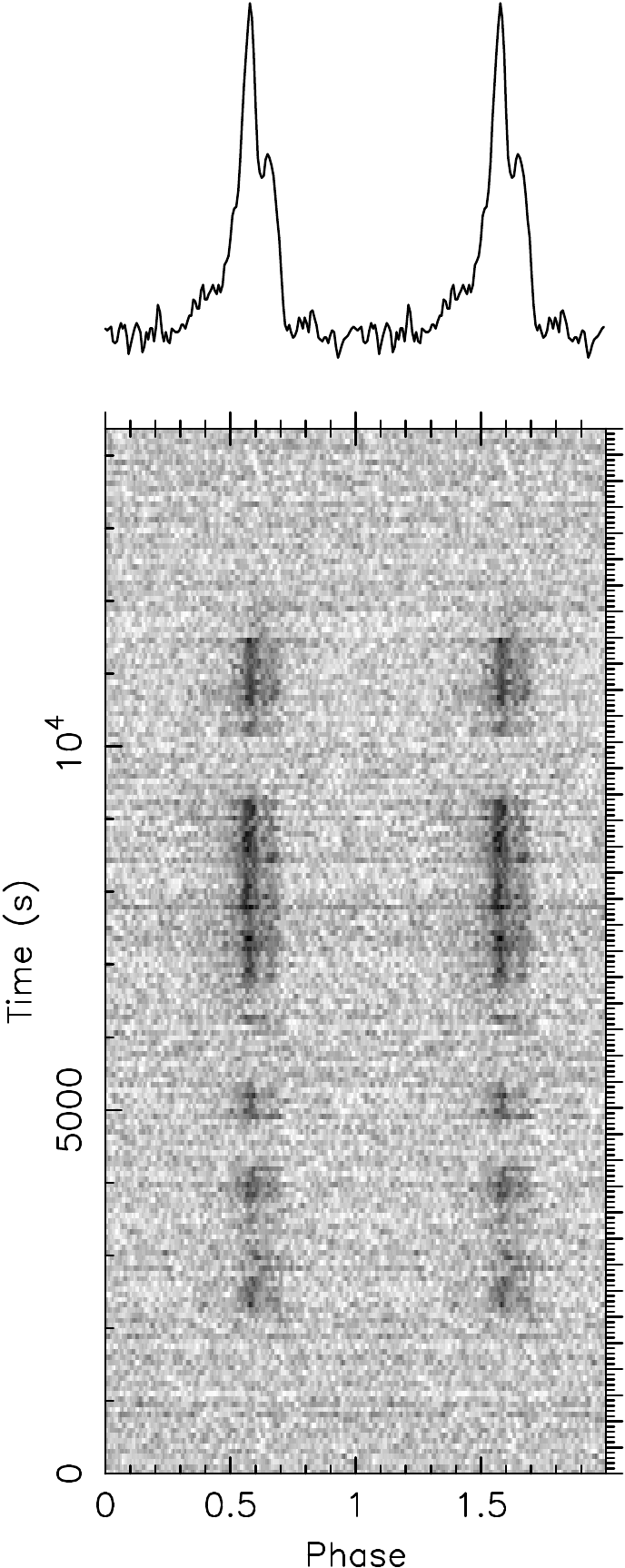}
    \qquad
	\includegraphics[width=0.44\columnwidth]{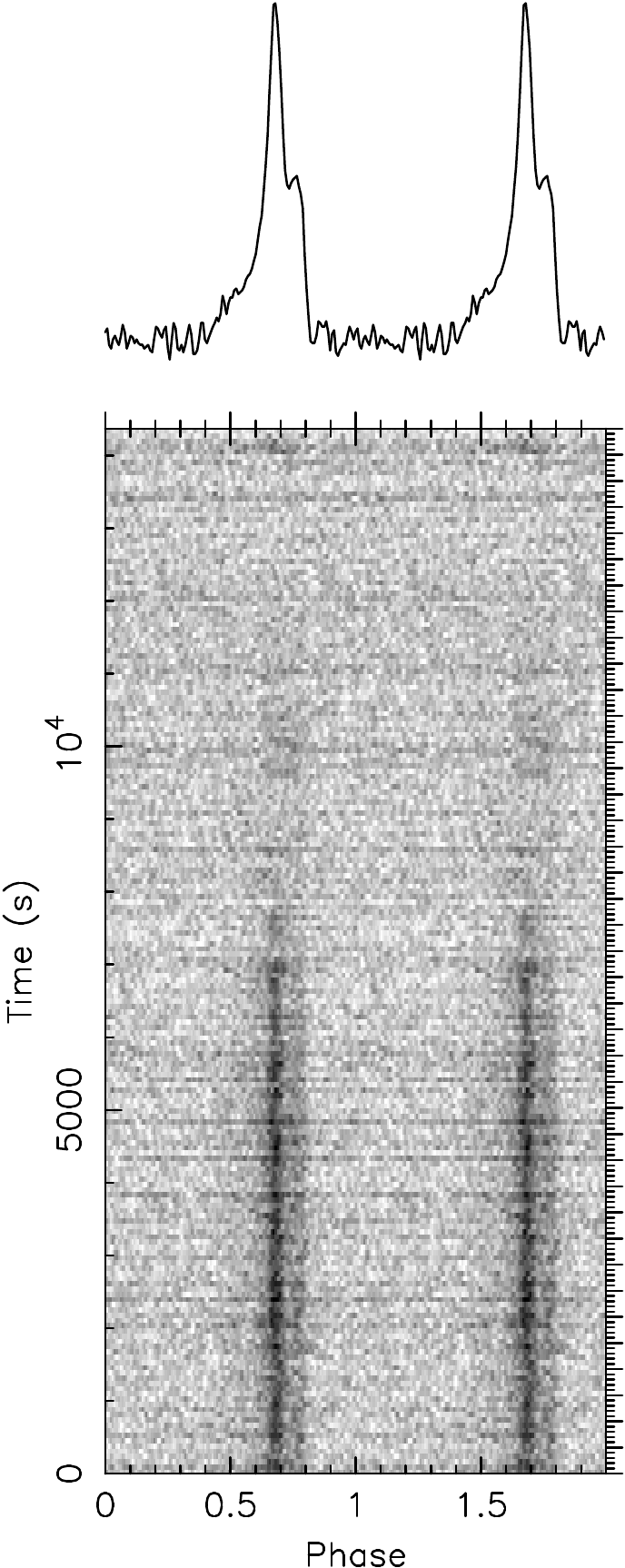}
  	\caption{ Intensity as a function of time and spin phase for the two strongest detections of the eclipsing redback NGC 6624F, obtained from the two 4-h-long TRAPUM observations of the cluster (left: obs. id 10L-trapum; right: obs. id 12L-trapum)}
  	\label{fig:NGC6624F_eclipses}
\end{figure}

\subsubsection{The orbit and eclipses of NGC 6624F}

NGC 6624F (or PSR J1823$-$3021F) is a 4.85-ms eclipsing binary pulsar discovered by \citet{Begin2006}. A first attempt on characterizing the orbit was made by \citet{Lynch+2012}, who estimated an orbital period of $\Pb = 0.8827$ d and a projected semi-major axis of $\xp \simeq 4.4$ lt-s, assuming a circular orbit. These estimates, however, were based on only three, fairly sparse in time detections, and hence not very robust. The many more, and much more closely-spaced MeerKAT detections have allowed us to obtain more precise and reliable measurements of the orbital parameters of NGC~6624F. We did so by starting with the Period-Acceleration diagram on the early observations, then fitting $\Pobs(t)$ from the dense orbital campaign observations, and finally fitting the ToAs obtained from all the detections.  We find that the actual orbital period of NGC 6624F is $\Pb=0.220682877(4)$~d, exactly one-fourth of the value reported by \citet{Lynch+2012}. The authors likely ended up overestimating $\Pb$ because of the sparsity of the three detections that they used.
The projected semi-major axis that we measure is $\xp=0.285389(5)$~lt-s, also much lower than the previously estimated value. The associated time of passage at the ascending node is $\Tasc = 58736.794401(3)$.
These orbital parameters imply a minimum companion mass $\Mc > 0.105$~\msun, hence NGC 6624F belongs to the class of redbacks.
As is common for many redbacks, NGC~6624F is always eclipsed for a large fraction ($\sim40$\%) of the orbit, around the pulsar's superior conjunction (in the range $\phib \sim 0.1-0.5$). On one occasion (i.e. the first TRAPUM observation, obs. id ``10L-trapum'', see Figure~\ref{fig:NGC6624F_eclipses}), the pulsar showed additional, short-lived eclipse events at orbital phases far from the superior conjunction.
None of the eclipses observed are accompanied by measurable delays of the pulses. Rather, the radio signal is attenuated fairly abruptly, indicating that the absorbing plasma is very dense, or has a small degree of ionization. Despite the many detections obtained with the MeerKAT observations, we were unable to derive an unambiguous phase connection for its ToAs across the $\sim11$ months spanned by our data. However, using the tiling of the TRAPUM observations, we could localize it at about 0.4 arcmin north of the cluster center (see next Section).

\begin{table}
\caption{List of the NGC 6624 pulsars without a known position and corresponding beams where they were detected with highest S/N in the first TRAPUM observation of the cluster (id 10L-trapum). $\theta_\perp$ is the angular distance between the beam boresight and the nominal cluster center.  }
\label{tab:NGC6624_localization}
\footnotesize
\centering
\renewcommand{\arraystretch}{1.0}
\begin{tabular}{ccccr}
\hline
\hline
\multicolumn{5}{c}{TRAPUM localization of the NGC 6624 pulsars}\\
\hline
Pulsar & Highest-S/N & Boresight RA & Boresight Dec   & $\theta_\perp$  \\
 & Beam (\#) & (hh:mm:ss)  & ($\degr:\arcmin:\arcsec$) & (arcsec) \\
\hline
D      & 004                        & 18:23:40.51 & $-$30:21:39.70 & 0                 \\
E      & 056                       & 18:23:38.93 & $-$30:22:22.30 & 47.25         \\
F      & 029                       & 18:23:40.67 & $-$30:21:14.20 & 25.58         \\
H      & 000                        & 18:23:40.48 & $-$30:21:39.90 & 0.44\\
\hline        
\end{tabular}
\end{table}
\subsubsection{Localization of five pulsars in NGC 6624}
\label{sec:localization}
\begin{figure*}
\centering
	\includegraphics[width=\textwidth]{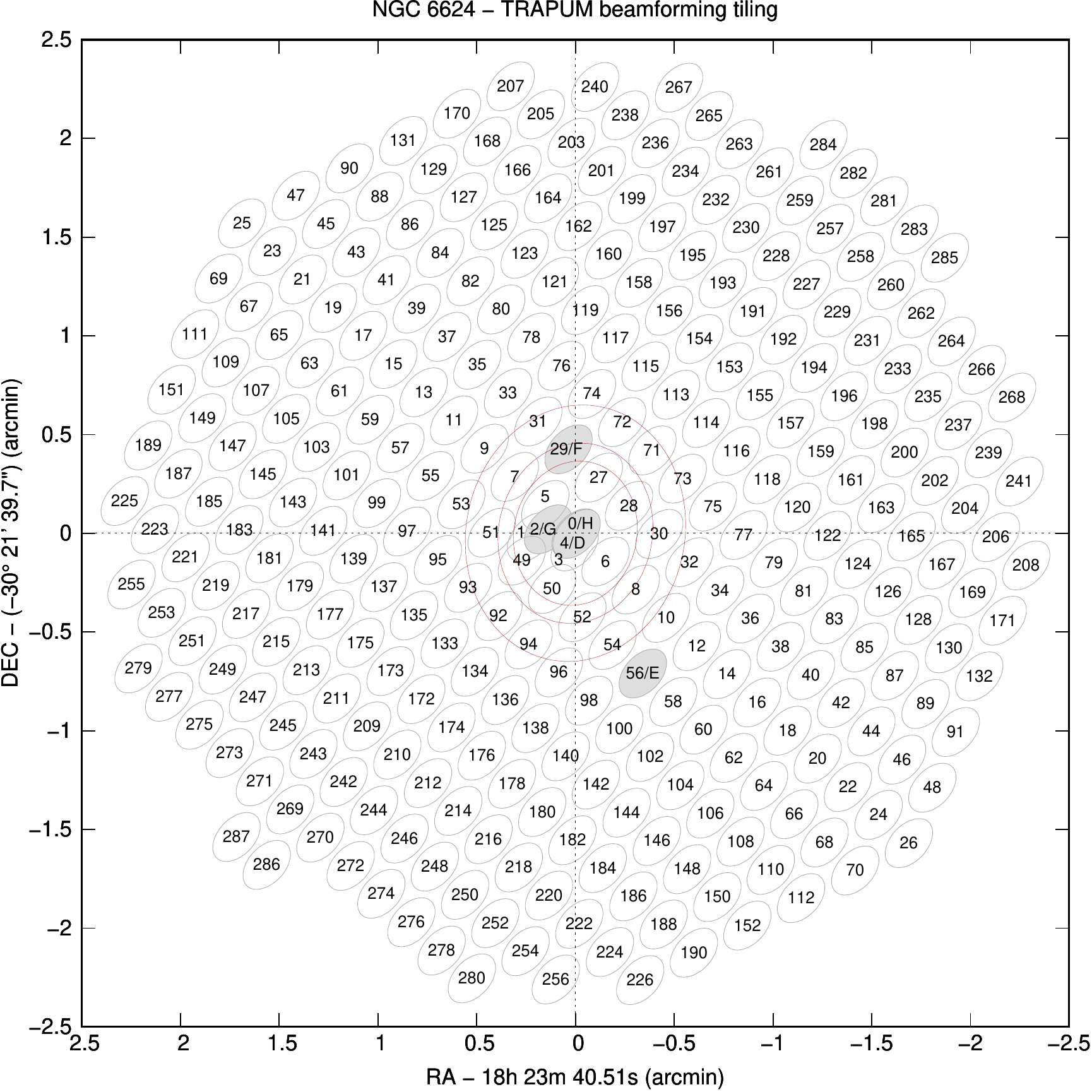}
  	\caption{Tiling of the first TRAPUM observation (obs. id 10L-trapum) of NGC 6624. 60 dishes were used to synthesize 288 beams, with an overlap fraction of 80\%, around the nominal center position of the cluster. The light-grey ellipses show the contours of the TRAPUM beams at their 84\% level of the power of their boresights, at 1284 MHz. We highlight in solid grey those beams where some of the pulsars show up with the highest S/N: the name of such pulsars are shown next to the beam number. the red ellipses show, for comparison, the L-band beam (at half power for top, center and bottom of the band) of one of the MeerTime observations of NGC 6624, which used only the 42 antennas of the 1-km core of the array.}
  	\label{fig:NGC6624_trapum_beams}
\end{figure*}

We took advantage of the tiling capabilities offered by the TRAPUM acquisition system to localize the pulsars currently lacking a phase-connected timing solution in NGC~6624 (pulsars D, E, F and H), as well as to validate the timing position derived for the new eccentric binary MSP (pulsar G). With the observing setup used for the two TRAPUM observations of this cluster (obs. ids 10L-trapum and 12L-trapum) reported in Section \ref{sec:observations}, the individual beams making up the tiling have elliptic shapes with semi-axes in the range of $\sim8-11$~arcsec at half power, with the exact values depending on the elevation and other factors. We assume these values as the maximum uncertainty for the positions that we derive for these pulsars. The results are summarized in Table \ref{tab:NGC6624_localization}, where we list the beams in which each pulsar was detected with highest S/N, and their boresight coordinates. The same is shown graphically in Figure \ref{fig:NGC6624_trapum_beams}, where the beams with localized pulsars are higlighted in grey. As can be seen, pulsars D and H are found to be the closest to the cluster center. The eclipsing redback F and the isolated pulsar E, are found at the much larger distances of $\sim0.42$~arcmin and $\sim0.79$ arcmin from the cluster center, respectively, putting the former near the edge of, and the latter well outside, the half-power of the tied-array beam of the MeerTime observations (which used only the 1-km core antennas) of NGC 6624. This explains why NGC~6624F and E are detected with significantly lower S/N in the MeerTime data as compared to the TRAPUM data. Finally, the eccentric binary pulsar NGC 6624G showed up with highest S/N in beam \#002. This had boresight coordinates of $\alpha=18\h23\m41\s.150$ and $\delta=-30\degr21\arcmin38\arcsec.50$. This is just $\sim 6.2$ arcsec away from, and therefore consistent with, the corresponding timing-derived position of G.

\begin{figure}
\centering \includegraphics[width=0.44\columnwidth]{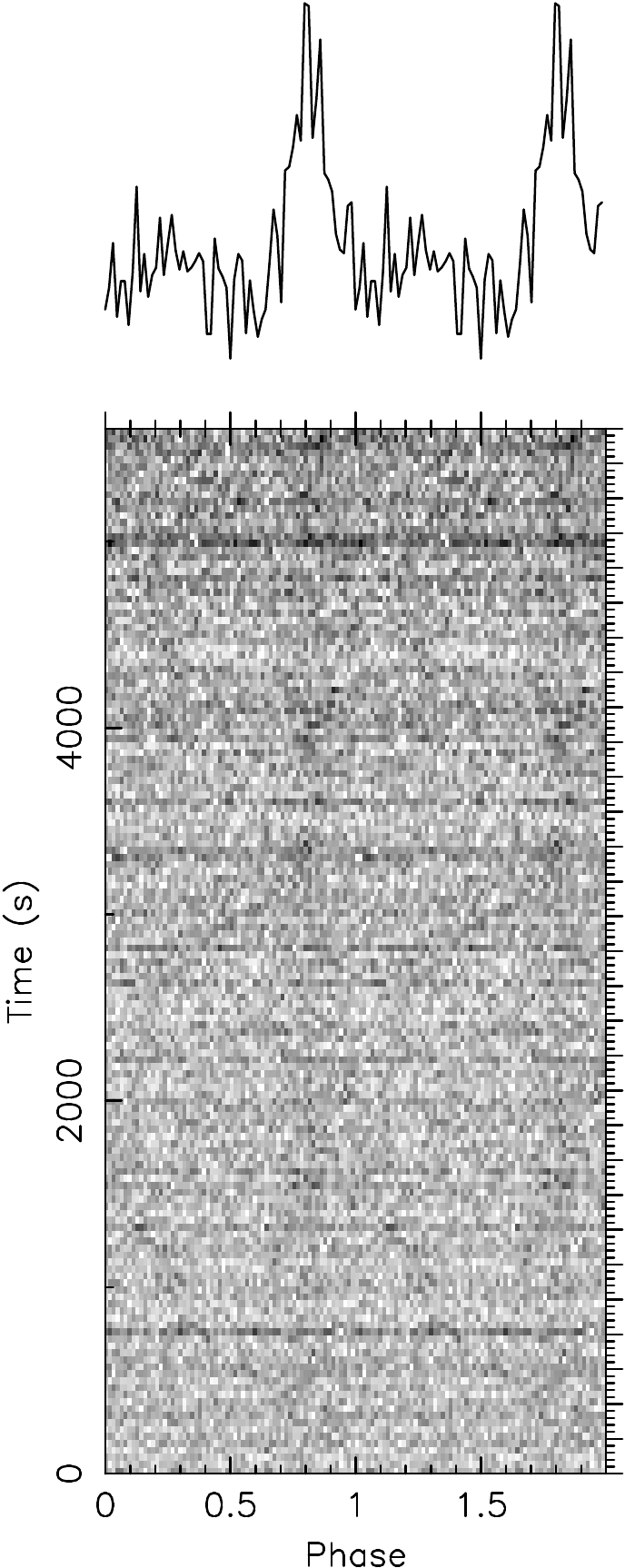}
	\qquad
    \includegraphics[width=0.44\columnwidth]{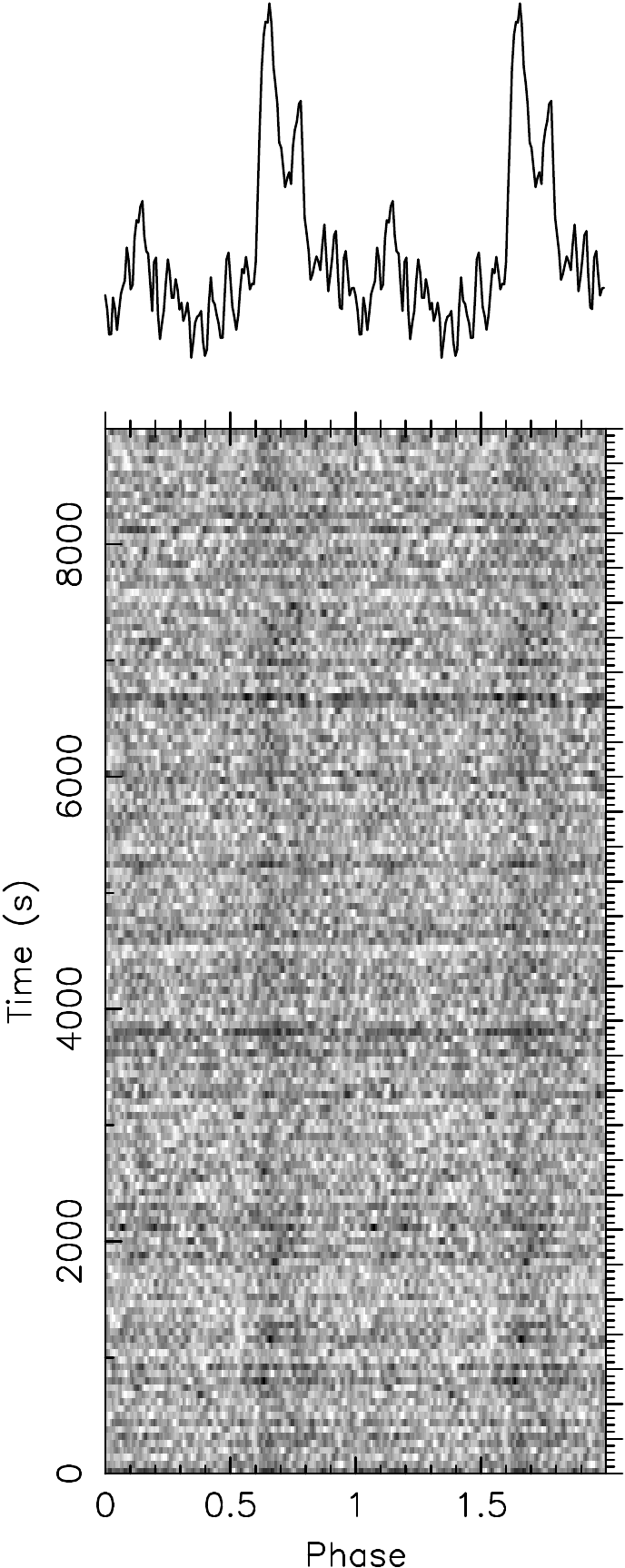}

  	\caption{Comparison between the two best detections of M28I obtained in the L band with the GBT on 2015 April 25 by \citet{Prager2017} (left), and with MeerKAT on 2019 July 19 (right). An interpulse is evident in the MeerKAT detection, whereas it is much less obvious in the former. }
  	\label{fig:M28I_01L_detection}
\end{figure}

\subsubsection{L-band detection of the transitional pulsar M28I}
M28I (also known as PSR J1824$-$2452I) is a redback system in M28 beloging to the rare class of \emph{transitional} MSPs (see \citealt{Jaodand+2018,Papitto_deMartino2020} for recent reviews). Discovered as a radio pulsar by \citet{Begin2006} with the GBT in the S band (central frequency of 1.95 GHz), it is the first system to be seen swinging between an accreting X-ray millisecond pulsar state, and a radio MSP state \citep{Papitto+2013,Linares+2014}, ultimately confirming the recycling scenario. In the context of our GC census, we have clearly detected M28I in the first  (obs id. 01L, dated 2019 July 19) of the two L-band MeerKAT observations of the host cluster made so far (Figure \ref{fig:M28I_01L_detection}). Hence, we can assert that on 2019 July 19, the pulsar was in a rotation-powered state. In the second MeerKAT observation of M28 (obs id. 02L, dated 2020 February 4), however, we were unable to re-detect the pulsar. This is not surprising since the pulsar is known to have a very erratic behaviour: with the GBT, M28I is detected in less than 20 per cent of the observations (S. Ransom, private comm.), and always at orbital phases in the range $\phib \sim 0.4-0.9$. What is more interesting and worthy of attention is the marginal detection (${\rm S/N}=6.6$) of an interpulse in the MeerKAT observation, which is not seen in the highest-S/N L-band  detection obtained with the GBT (Figure \ref{fig:M28I_01L_detection}) in 2015. This could potentially have arisen from a rearrangement of the pulsar's magnetosphere after some short accretion episode. However, the low S/N of the GBT detection and lack of additional information currently prevents us from drawing any firm conclusions.


\section{DISCUSSION AND OUTLOOK}
\label{sec:discussion}

\subsection{Summary of the findings}

In this paper we have presented the discovery and early follow-up of eight new MSPs in a set of six GCs, out of a total of nine clusters targeted by this survey.

Overall, absent any theoretical interpretation, we find that our findings exacerbate the observed differences between the pulsar populations of different GCs. 
In core-collapsed clusters, where isolated pulsars were dominant, three out of four new discoveries (NGC 6522D, NGC 6624H and NGC 6752F) were isolated MSPs. In the remaining 
GCs we have only discovered binaries. 

This is very much in line with the expectations of \cite{Verbunt_Freire_14}.
For clusters with high interaction rate, but a low interaction rate per star, like 47 Tuc, the vast majority of the many LMXBs that form evolve undisturbed, resulting in a population of fully-recycled MSPs (all pulsars in 47~Tuc have spin periods between 2 and 8 ms)
in circular orbits with WD companions, many compact ``spider'' systems with very light degenerate or semi-degenerate
companions \citep{Roberts2013}, and some isolated pulsars; i.e., a population that is nearly identical
to the Galactic population of pulsars with these spin periods.
With their characteristics, the newly discovered 47~Tuc~ac and 47~Tuc~ad fit very well within the previously known pulsar population
in 47 Tuc, which already had five black widows (47~Tuc~I, J, O, P, R) and two redbacks (47~Tuc~V and W).
The new binary in M62 also fits well with the other six binaries in that cluster, which, like 47~Tuc, also has a mix of MSP-WD systems, redbacks and black widows (but thus far no isolated pulsars).

For core-collapsed GCs, which have a
large stellar interaction rate per star, most of the binary pulsars (or even LMXBs) have dissociated because 
of the high rate of interactions. This means that the pulsar populations in such clusters are dominated by isolated pulsars, and some only partially recycled and thus apparently young (like NGC 6624A, B and C), with some of the latter being quite slow (NGC 6624B and C).
NGC 6522 had three isolated MSPs, now it has four, and no known binaries. NGC~6752 had four isolated MSP and one binary, now it has one more isolated MSP. NGC~6624 had five isolated pulsars and one binary; we discovered one extra isolated pulsar (NGC~6624H) and one binary, NGC~6624G, the only such system we discovered in a core-collapsed cluster. However, and as we have seen, even this system is unusual: it is likely a secondary exchange encounter.
As mentioned in the section on target selection (Section \ref{sec:selection}), finding this sort of systems was precisely the reason why we targeted core-collapsed clusters. The discovery of NGC~6624G confirms that searching core-collapsed GCs is a fruitful strategy.

The GC Ter~5 has intermediate characteristics: although not a core-collapsed cluster, it has a relatively high encounter rate per binary, thanks to its very dense core. It has a few binaries that are very likely secondary exchange encounters. The new binary system discovered there, Ter 5 an, is similar to other binary systems found in that cluster, like Ter 5 N and W: their eccentricities are probably too low for them to have originated in a secondary exchange encounter (they are more likely caused by close flybys of other stars). However, the companion masses are somewhat large compared to the masses of the He WD companions to fast-spinning MSPs we find in the Galactic disk.

\subsection{Follow-up of the discoveries}

To fully exploit the scientific potential of the discoveries, follow-up observations are highly desirable. For this reason, we have set up a GC pulsar follow-up programme, whose main goals are the confirmation and/or the timing of the new discoveries, which makes use of MeerKAT itself as well as other radio telescopes with comparable sensitivity or complementary characteristics. For example, the Parkes radio telescope, in combination with its Ultra-Wideband Low receiver \citep{Hobbs+2020}, may be an asset for confirming discoveries in clusters that are affected by strong scintillation. 100-m class radio telescopes such as the GBT and the Effelsberg telescope will also be important for the long-term timing of some of the discoveries.
As of December 2020, we already have two proposals approved at two different telescopes: a candidate-confirmation programme using Parkes, and a follow-up timing of NGC 6624G using the GBT with its S-band and L-band receivers.
For the faintest discoveries presented here, such as NGC 6624H, and NGC 6522D, the use of MeerKAT for their follow-up is probably necessary. However, their isolated nature will make it possible to derive their timing solutions conducting observations with a relatively low cadence.

\subsection{The future TRAPUM GC pulsar survey}
Besides discovering new pulsars, the GC pulsar census served as a testbed and paved the way for the upcoming, fully-fledged TRAPUM GC pulsar survey. The latter will take advantage of the whole MeerKAT array, using a minimum of 56 and a maximum of 64 antennas in each observation. This is opposed to the census survey observations, which exploited only the 1-km core of the array, with 38--42 antennas typically used. These numbers imply that the nominal combined telescope gain for the TRAPUM GC survey will be between 33 and 68 per cent higher than that of the census observations, reaching 2.8~K\,Jy$^{-1}$ when using all the 64 antennas. Such an increase in raw sensitivity will be even more pronounced when searching in regions at angular distances $\gtrsim 0.5$~arcmin from the cluster center, i.e. outside the typical half-power radius of the MeerTime tied-array beam at L band for the census observations. This is particularly relevant for nearby and extended clusters, such as NGC 6397, 47~Tuc or $\omega$~Centauri, to mention a few.

\section{Conclusions}
\label{sec:conclusions}
We have used the MeerKAT radio telescope to conduct a first census of pulsars in nine selected globular clusters. As part of this programme, we have carried out a deep search for new pulsars using the 1-km core of the array at L-band, as well as a few follow-up observations with different configurations, two of which exploited the full array and the tiling capabilities of the TRAPUM backend. We have discovered eight new MSPs in six different clusters. Five of these new pulsars are part of binary systems. One of them, NGC 6624G, is one of the rare MSPs in highly eccentric orbits that resulted from exchange encounters occurring in the extremely dense globular cluster environments. Early timing of this pulsar indicates that the system is massive, with a 69 per cent probability for the NS to exceed 2~\msun. Similar (although more uncertain) conclusions apply to another of the discovered binary pulsars, Ter~5~an. Further follow-up timing observations are necessary to constrain the masses of these two systems with higher precision.
The same census data was used to better characterize some of the previously known pulsars.
These early results demonstrate the exceptional capabilities of the MeerKAT telescope both for the searching and timing of GC pulsars, especially (but not exclusively) for southern sky clusters. We have also shown how the science of pulsars in GCs can greatly benefit from the synergy between the MeerTime and TRAPUM projects, as a result of their complementary goals and characteristics. 
Notwithstanding, archival data taken at other telescopes have once again proven to be essential for the confirmation and/or characterization of many of the new pulsars, allowing for an instantaneous multi-year timing baseline for a few of them. This is very peculiar to globular cluster pulsar observations, and shows how important it is to keep archival search-mode data. 
The eight new MSPs presented in this work are the first of several GC pulsar discoveries so far made by MeerKAT \footnote{See \url{http://www.trapum.org/discoveries.html} for an up-to-date list.}. They represent some of the early examples of a new wave of discoveries of GC pulsars, which have, at the time of writing, reached a total of 221.

\section*{Acknowledgements}
The MeerKAT telescope is operated by the South African Radio Astronomy Observatory, which is a facility of the National Research Foundation, an agency of the Department of Science and Innovation.
PTUSE was developed with support from  the Australian SKA Office and Swinburne University of Technology.
The Parkes radio telescope is funded by the Commonwealth of Australia for operation as a National Facility managed by CSIRO. 
The National Radio Astronomy Observatory is a facility of the National Science Foundation operated under cooperative agreement by Associated Universities, Inc. 
The Green Bank Observatory is a facility of the National Science Foundation operated under cooperative agreement by Associated Universities, Inc.
This work was partly performed on the OzSTAR national facility at Swinburne University of Technology. The OzSTAR program receives funding in part from the Astronomy National Collaborative Research Infrastructure Strategy (NCRIS) allocation provided by the Australian Government.
The authors also acknowledge MPIfR funding to contribute to MeerTime infrastructure.
MBu and AR thank the Autonomous Region of Sardinia (RAS) for funding computing resources that were used for this work (Regional Law 7 August 2007 n. 7, year 2015, ``Highly qualified human capital''; project CRP 18 ``General relativity tests with the Sardinia Radio Telescope''; P.I. Marta Burgay)."
This research was funded partially by the Australian Government through the Australian Research Council, grants CE170100004 (OzGrav) and FL150100148.
AC, APo, AR and MBu gratefully acknowledge financial support by the research grant ``iPeska'' (P.I. Andrea Possenti) funded under the INAF national call Prin-SKA/CTA approved with the Presidential Decree 70/2016.
APo, AR, MBu acknowledge the support from the Ministero degli Affari Esteri e della Cooperazione Internazionale - Direzione Generale per la Promozione del Sistema Paese - Progetto di Grande Rilevanza ZA18GR02.
SMR is a CIFAR Fellow and is supported by the NSF Physics Frontiers Center award 1430284.
BWS acknowledges funding from the European Research Council (ERC) under the European Union’s Horizon 2020 research and innovation programme (grant agreement No. 694745).
FA, MK, and PVP gratefully acknowledges support from ERC Synergy Grant ``BlackHoleCam'' Grant Agreement Number 610058.
AR, TG, PCCF, VVK, MK, FA, EDB, PVP, DJC, RK, and APa acknowledge continuing valuable support from the Max-Planck Society.
LVC acknowledges financial support from the Dean’s Doctoral Scholar Award from the University of Manchester.
RMS acknowledges support through the Australian Research Council (ARC) Future Fellowship FT190100155.
MBa and RMS acknowledge support through ARC grant CE170100004.
AK acknowledges support from the UK Science and Technology Facilities Council, grant ST/S000488.
We thank Ryan Lynch for providing GBT data that was important to validate and improve some of the timing results presented here.
We are grateful to the anonymous referee for useful comments, which helped improve the manuscript.

\section*{Data Availability}
The data underlying this article will be shared on reasonable request to the MeerTime and TRAPUM collaborations.




\bibliographystyle{mnras}
\bibliography{MeerKAT_GC_census} 


%
%
%


\bsp	
\label{lastpage}
\end{document}